\documentclass[reprint,showpacs,preprintnumbers,amsmath,amssymb]{revtex4-1}

\pdfoutput=1
\usepackage[pdftex]{graphicx}
\usepackage{color}
\definecolor{darkred}{rgb}{0.4,0.0,0.0}
\definecolor{darkgreen}{rgb}{0.0,0.4,0.0}
\definecolor{darkblue}{rgb}{0.0,0.0,0.4}
\usepackage[bookmarks,linktocpage,colorlinks,linkcolor = darkred,urlcolor  = darkblue,citecolor = darkgreen,filecolor=cyano]{hyperref}
\usepackage{braket}
\usepackage[T1]{fontenc}
\usepackage{amsmath}
\usepackage{amsthm}
\usepackage{amssymb}
\usepackage{url}
\usepackage[utf8]{inputenc}
\usepackage[english]{babel}
\usepackage{dcolumn}% Align table columns on decimal point
\usepackage{bm}% bold math
\usepackage{subfig}
\usepackage{float}
\usepackage{url} 
\usepackage{grffile}
\usepackage{xcolor}
\usepackage{booktabs}
\usepackage{verbatim}

\newcommand{\Tr}{\mbox{\rm Tr\,}}

\newcommand{\be}{\begin{equation}}
\newcommand{\ee}{\end{equation}}
\newcommand{\bea}{\begin{eqnarray}}
\newcommand{\eea}{\end{eqnarray}}
\newcommand{\non}{\nonumber}
\newcommand{\bie}{\begin{small} \begin{itemize}}

\newcommand{\eie}{\end{itemize} \end{small}}
\usepackage{cleveref}
\crefname{section}{Section}{Sections}
\Crefname{section}{Section}{Sections}
\crefname{equation}{Eq.}{Eqs.}
\Crefname{equation}{Eq.}{Eqs.}
\crefname{figure}{Fig.}{Figs.}
\Crefname{figure}{Fig.}{Figs.}
\crefname{table}{Table}{Tables}
\Crefname{table}{Table}{Tables}

\graphicspath{{lat_fig/}}

\usepackage{ragged2e}

\begin{document}
%%%%%%%%%%%%%%%%%%%%%%%%%%%%%%%%%%%%%%%%%%%%%%%%%%%%%%%%%%%%%%%%%%%%%%%%%%%%%
%
\selectlanguage{english}
%----------------------------------------------------------------------------
\title{%
%\textcolor{cyano}{\huge \em PRELIMINARY DRAFT} \\
Spectrum of very excited $\Sigma_g^+$ flux tubes in SU(3) gauge theory}
%----------------------------------------------------------------------------
\author{P. Bicudo}
\email{bicudo@tecnico.ulisboa.pt}
\author{N. Cardoso}
\email{nuno.cardoso@tecnico.ulisboa.pt}
\author{A. Sharifian}
\email{alireza.sharifian@tecnico.ulisboa.pt}
\affiliation{CeFEMA, Departamento de F\'{\i}sica, Instituto Superior T\'{e}cnico
(Universidade T\'{e}cnica de Lisboa),
Av. Rovisco Pais, 1049-001 Lisboa, Portugal}
%----------------------------------------------------------------------------
\begin{abstract}
Spectra with full towers of levels are expected due to the quantization of the string vibrations,
however different theoretical models exist for the excitation spectra.
First principle computations are important to test the different models and to search for novel phenomena,
but so far only a few excited states of QCD flux tubes have been studied with pure gauge SU(3) lattice QCD in 3+1 dimensions.
We thus aim to study a spectrum of flux tubes with static quark and antiquark sources up to a significant number of excitations. 
We specialize on the spectrum of the most symmetric case, namely $\Sigma_g^+$, where up to two levels are already published in the literature. To achieve the highest possible excitation level, we construct a large  set of operators with the correct symmetry, solve the generalized eigenvalue problem and compare the results of different lattice QCD gauge actions with different lattice spacings and anisotropies. 
\end{abstract}
%----------------------------------------------------------------------------
\maketitle
%----------------------------------------------------------------------------

%SSSSSSSSSSSSSSSSSSSSSSSSSSSSSSSSSSSSSSSSSSSSSSSSSSSSSSSSSSSSSSSSSSSSSSSS
%SSSSSSSSSSSSSSSSSSSSSSSSSSSSSSSSSSSSSSSSSSSSSSSSSSSSSSSSSSSSSSSSSSSSSSSS
%SSSSSSSSSSSSSSSSSSSSSSSSSSSSSSSSSSSSSSSSSSSSSSSSSSSSSSSSSSSSSSSSSSSSSSSS
\section{Introduction \label{sec:intro}}

Understanding the confinement of colour remains a main theoretical problem of modern physics. Its solution could also open the door to other unsolved theoretical problems. 
An important evidence of confinement, where we may search for relevant details to understand it, is in the QCD flux tubes \cite{Wilson:1974sk} 
\textcolor{black}{
computed in lattice QCD. 
Experimentally, flux tubes are suggested by the Regge trajectories 
\cite{Bicudo:2007wt,Bicudo:2009hm}
in the hadron spectrum. 
String models can account for Regge trajectories, and they also imply the linear confining quark-antiquark potential  similar to the one used in quark models
\cite{Godfrey:1985xj,Isgur:1978xj}. 
}

Presently, an approximate understanding of flux tubes is quite developed, for example with string models.
The dominant behaviour of the  string-like flux tubes has a single scale: the string tension $\sigma$. 
The main analytical string model utilised in the literature to explain the behaviour of the QCD flux tubes is the Nambu-Goto bosonic \cite{Nambu:1978bd,Goto:1971ce} string model \cite{Aharony:2009gg}. It assumes infinitely thin strings, with transverse quantum fluctuations only. The quantum fluctuations predict not only a finite profile width of the groundstate flux tube, increasing with distance \cite{Gliozzi:2010zv}, but also an infinite tower of quantum excitations \cite{Alvarez:1981kc,Arvis:1983fp}. However, the string-like behaviour obscures the details of confinement and of other possible hadronic phenomena.

Clearly, at short quark-antiquark distances, the flux tube deviates from the string model. The Nambu-Goto model in four space-time dimensions has an imaginary tachyon \cite{Arvis:1983fp} at short distances for the groundstate, whereas the QCD flux tube has a real Coulomb potential 
\cite{Arvis:1983fp}. At really short distances, lattice QCD has recently shown the potential becomes dominated by perturbative QCD 
\cite{Karbstein:2014bsa}.

Another instance where the groundstate flux tube deviates from the string model is in the flux tube profile. Recently, our lattice QCD collaboration  PtQCD 
\cite{PtQCD} 
studied the zero temperature groundstate flux tube of pure gauge QCD, and found evidence for a penetration length $\lambda$ 
\cite{Cardoso:2013lla}, 
as a second scale other than the string tension $\sigma$, contributing to the colour fields density profile of the flux tube. This intrinsic width of the flux tube may be the reason why the QCD flux tube is stable in four dimensions, unlike the Nambu-Goto string which is rotational invariant only in 26 dimensions.

There is also an ongoing puzzle in the excited spectrum of mesons, as reported \cite{Bugg:2004xu} in measurements by the Cristal Barrel detectors \cite{Aker:1992ny}: the Regge slope for radial excitations is similar to the one for angular excitations, and this cannot be explained with a quark model \cite{Bicudo:2007wt}. A large degeneracy, larger than the chiral restoration symmetry \cite{Glozman:2007ek}, has been analysed \cite{Afonin:2006wt,Afonin:2007jd,Glozman:2012fj,Catillo:2018cyv}. Possibly there is a new principal quantum number \cite{Bicudo:2009hm}. 
\textcolor{black}{
Notice a principal quantum number is already present in the  Nambu-Goto spectrum.
}

\textcolor{black}{
Besides, an infinite tower \cite{Semay:2008nq} of these excitations can also be obtained with constituent gluon models, in the denominated hybrid three body quark-gluon-antiquark systems \cite{Abreu:2005uw,Buisseret:2006wc}. 
This may possibly be explained by \cite{Bicudo:2009hm}
}
a constituent gluon with an effective mass
\cite{Cornwall:1981zr,Oliveira:2010xc}, and an effective
quark gluon-gluon potential
\cite{Bicudo:2007xp,Cardoso:2009kz}.

\textcolor{black}{
Moreover, in there is evidence for another particle, the string worldsheet axion \cite{Dubovsky:2013gi}, in the lattice gauge theory spectra of closed strings.
}

\textcolor{black}{
Thus precise first principle theoretical studies are necessary to go beyond the string models and clarify the details of QCD flux tubes, or hybrid.
In lattice QCD it is straightforward to study groundstate 
open
}
 flux tubes, using the technique of Wilson loops  \cite{Wilson:1974sk}, the simplest gauge invariant correlations in lattice QCD. 

\textcolor{black}{
There are two classes of flux tubes, the closed flux tubes and the open ones.
In lattice QCD, they are respectively studied with torelons and Wilson loops. There has been a serious effort of extracting higher excitations
in the closed flux-tube channel
\cite{Athenodorou:2010cs,Lucini:2012gg}. 
The spectrum of the closed flux-tube can be partially approximated by the Nambu-Goto  model.
}

In this paper we study the excited 
\textcolor{black}{
open
}
 flux tubes in lattice QCD, in particular we opt to address the puzzling sector of radial excitations.
Some excited states have indeed been observed by lattice QCD computations \cite{Campbell:1987nv,Perantonis:1990dy,Lacock:1996ny,Lacock:1996vy,Juge:1999ie,Juge:2002br,Reisinger:2017btr}, so far compatible with the string dominance of the QCD flux tube. Moreover the structure of excited flux tubes started to be studied as well \cite{Mueller:2018fkg,Bicudo:2018jbb,Mueller:2019mkh}, where some evidence for a constituent gluon is present in some states. However only the first excitations have so far been studied.

The spectrum of radial excitations of the groundstate is the $\Sigma_g^+$ spectrum. We now review the symmetry group of our flux tubes. With two static sources, it is equivalent to the one of the molecular orbitals of homonuclear diatomic molecules. It is the point group denominated $D_{\infty h}$. We thus utilise the standard quantum number notation of molecular physics, already adopted in the previous studies of QCD flux tube excitations \cite{Campbell:1987nv,Perantonis:1990dy,Lacock:1996ny,Lacock:1996vy,Juge:1999ie,Juge:2002br,Reisinger:2017btr,Bicudo:2018jbb,Mueller:2019mkh}. 
$D_{\infty h}$ has three symmetry sub-groups, and they determine three quantum numbers.

The two-dimensional rotation about the charge axis corresponds to the quantum angular number, projected in the unit vector of the charge axis  $\Lambda = \left| {\bf J}_g\!\cdot \hat e_z \right| $. The capital Greek
letters $\Sigma, \Pi, \Delta, \Phi, \Gamma \dots$ indicate as usually states
with $\Lambda=0,1,2,3,4 \dots$, respectively. The notation is reminiscent of the $s, \, p, \, d \cdots $ waves in atomic physics.  In the case of two-dimensional rotations there are only two projections ${\bf J}_g\!\cdot \hat e_z = \pm \Lambda$.

The permutation of the quark and the antiquark static charges is equivalent to a combined operations of
charge conjugation and spatial inversion about the origin. Its eigenvalue is denoted by
$\eta_{CP}$.  States with $\eta_{CP}=1 (-1)$ are denoted
by the subscripts $g$ ($u$), short notation for {\em gerade} ( {\em ungerade}).  

Moreover, there is a third quantum number, different from the phase corresponding to a two-dimensional p-wave. 
Due to the planar, and not three-dimensional, angular momentum there is an additional label for the s-wave
$\Sigma$ states only. $\Sigma$ states which
are even (odd) under the reflection about a plane containing the molecular
axis are denoted by a superscript $+$ $(-)$. 

With these quantum numbers, the energy levels of the flux tubes are labeled  as $\Sigma_g^+$, $\Sigma_g^-$, $\Sigma_u^+$, $\Sigma_u^-$,
$\Pi_u$, $\Pi_g$, $\Delta_g$, $\Delta_u \cdots$
The states we opt to study are most symmetric system ones, corresponding to the $\Sigma_g^+$ spectrum. This is the spectrum where up to two levels have so far been reported in the lattice QCD simulations literature \cite{Juge:1999ie,Juge:2002br}. 
For the other spectra only the groundstate level has been reported in the literature \cite{Campbell:1987nv,Perantonis:1990dy,Lacock:1996ny,Lacock:1996vy,Juge:1999ie,Juge:2002br,Reisinger:2017btr}.
In principle, the $\Sigma_g^+$ is the most amenable for a first study of very excited states.

In Section \ref{sec:framework} we thus detail the different state of the art lattice QCD techniques adequate to study the excited spectrum of the gluonic flux tubes  \cite{Campbell:1987nv,Perantonis:1990dy,Lacock:1996ny,Lacock:1996vy,Juge:1999ie,Juge:2002br} produced by a static quark-antiquark pair.
Working with pure gauge SU(3) fields discretised in a lattice, we utilise Wilson loops with a large basis of gluonic spacelike Wilson lines to include different excitations of the quark-antiquark flux tube. Moreover, we apply different smearing techniques, an improved action, and an anisotropic lattice to improve the signal over noise ratio.
We combine our operators, to block diagonalize our basis the angular momentum and parity quantum numbers of the $D_{\infty h}$ point group and project on $\Sigma_g^+$ states.
We numerically solve the correlation matrix via the Generalized Eigenvalue Problem and compute the corresponding effective masses.
We also discuss our computational efficiency. The number of gluonic operators combined with the space points where we compute the flux tube densities turn out to be very large, and we resort to computations in GPUs and to CUDA codes.

In Section \ref{sec:results} we show, for the excited levels where the signal is clear, the results of our computations for the spectrum. 
We evaluate how the different techniques improve the results.
Finally in Section \ref{sec:analysis}, we analyse our results for the first excitations of the flux tube and search for signals of novel phenomena beyond the Nambu-Goto string model; in Section \ref{sec:conclu} we conclude our work.

%SSSSSSSSSSSSSSSSSSSSSSSSSSSSSSSSSSSSSSSSSSSSSSSSSSSSSSSSSSSSSSSSSSSSSSSS
%SSSSSSSSSSSSSSSSSSSSSSSSSSSSSSSSSSSSSSSSSSSSSSSSSSSSSSSSSSSSSSSSSSSSSSSS
%SSSSSSSSSSSSSSSSSSSSSSSSSSSSSSSSSSSSSSSSSSSSSSSSSSSSSSSSSSSSSSSSSSSSSSSS
\section{Lattice QCD framework to compute the $\Sigma_g^+$ flux tubes \label{sec:framework}}

We discuss our strategy to compute as many radially excited states as we can for the $\Sigma_g^+$ flux tubes. We consider a large basis of operators, use the standard action and an improved ones, apply different smearing techniques in the configurations, and use different lattices with space-time anisotropy.

%
%--------------
\begin{figure}[t!]% no figure before 1st section
\begin{centering}
\includegraphics[width=0.95\columnwidth]{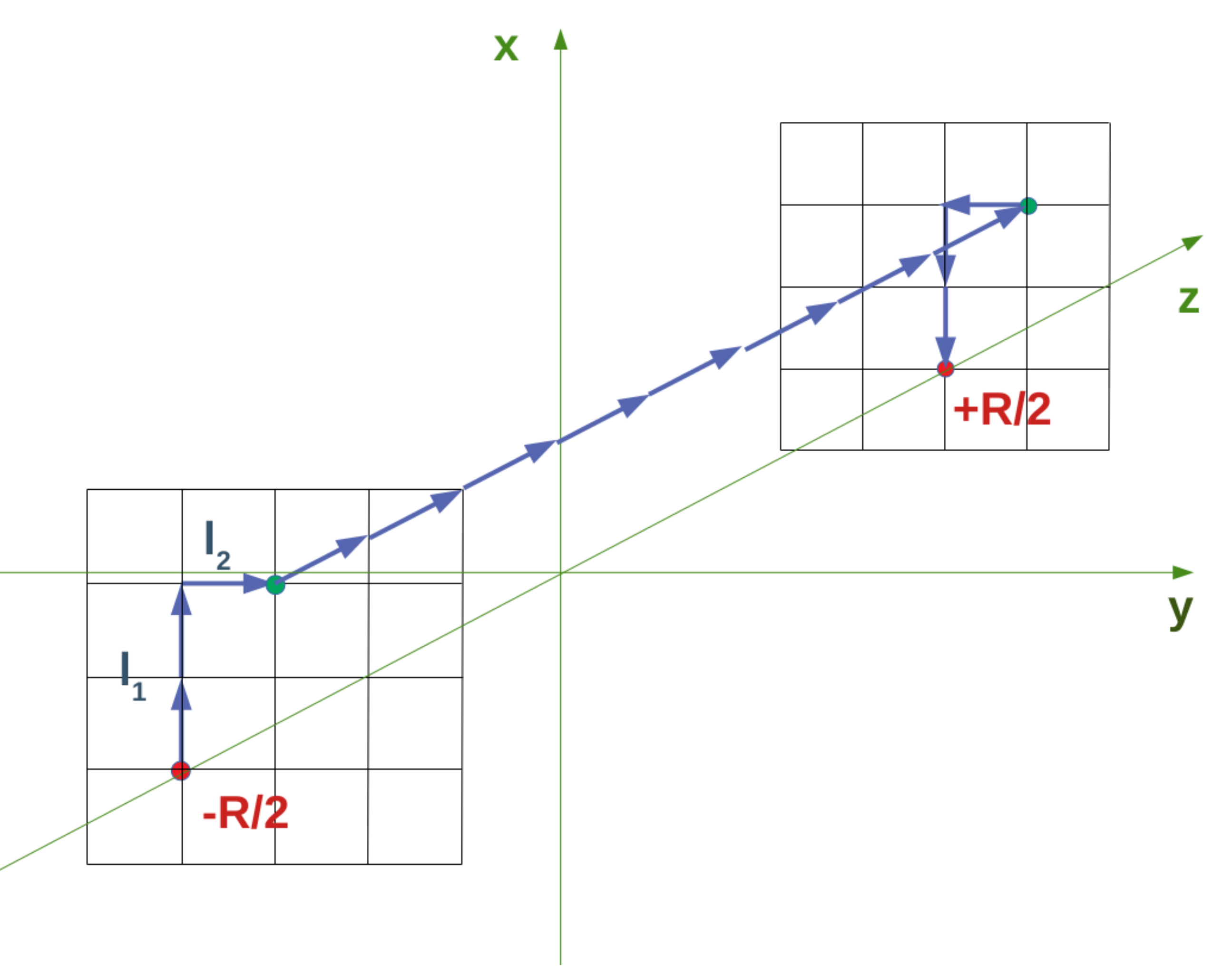}
\caption{ Example of a Wilson spacial path from the antiquark to the quark, to be used as one of the components of the flux tube operator
\textcolor{black}{
$O(2,1)$.
}
The $z$ axis is the charges axis and $xy$ is the perpendicular spatial plane.
\label{fig:exoperators}}
\par\end{centering}
\end{figure}

\subsection{Our operator basis  for the   $\Sigma_g^+$ excited states \label{sec:operator}}

Our gauge invariant operators are a collection of Wilson loops, a closed path of Wilson lines.  
Since we have static charges, our temporal Wilson lines are quite simple, they are straight lines.

We want our spatial operators to have the same symmetry as the flux tubes we are studying. Moreover, since we are using a correlation matrix, we want any linear combination of our operators to also have the same symmetry. Otherwise we could be producing unwanted states. The symmetry is provided by the spacial Wilson lines, who close the Wilson loop and make it gauge invariant.

In the study of the flux tubes, we utilise a basis of spacial Wilson line operators, with the necessary and sufficient operators to produce the $\Sigma_g^+$ spectrum and avoid producing the other spectra. 
As usual we choose our frame such that the charge axis is the $z$ axis, and the origin is set at the midpoint between the quark and the antiquark, with distance $R$. The $x$ and $y$ axis are in the two perpendicular directions.

%
%--------------
\begin{figure}[t!]% no figure before 1st section
\begin{centering}
\includegraphics[width=0.95\columnwidth]{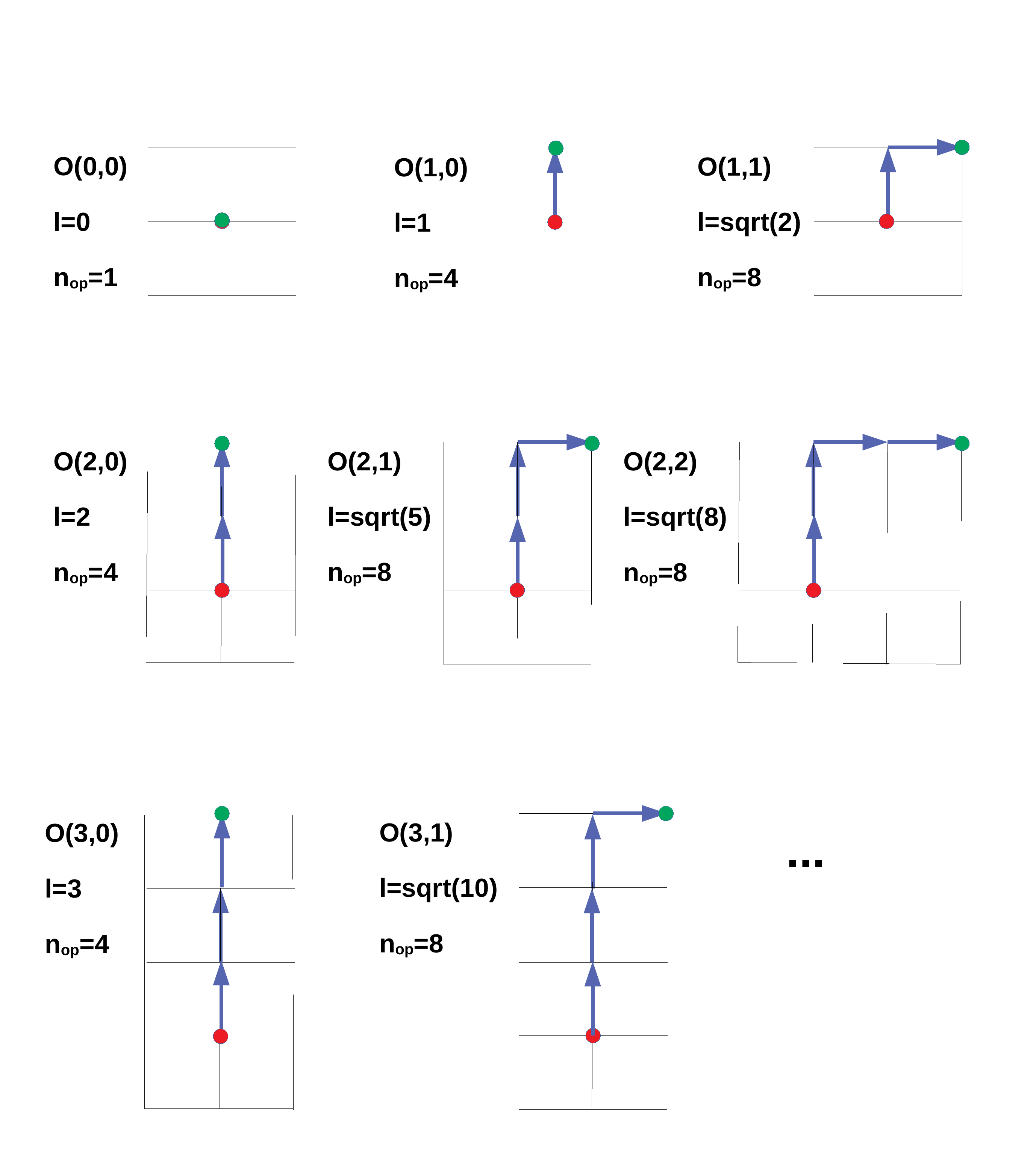}
\caption{Detail of the section, in the $xy$ plane perpendicular to the charge $z$ axis,  of Wilson spacial paths from the antiquark to the quark. These examples correspond to a sub-operator  used in the gauge field operators $O(l_1, l_2)$. The full operator has the symmetric sum of the corresponding set of $n_\text{op}$ operators, all at the same distance  $l= \sqrt{{l_1}^2+ {l_2}^2}$ from the charge $z$ axis.
\label{fig:operators1}}
\par\end{centering}
\end{figure}

We first must choose a basis of operators, consisting of a linear combination of a set of spatial
Wilson curves connecting the two static charges, with the same symmetries as those of $\Sigma_g^+$. Each curve we consider first departs from the antiquark charge in the $xy$ plane perpendicular to the charge. In Fig.  \ref{fig:operators1} we show different such curves. Then the path of the operator is continued with a straight Wilson line in the $z$ direction, and finally it is completed with an opposite curve in the $x,y$ plane to join the quark charge. 
The operators must be invariant for rotations of angles multiple of $\pi /2$ around the charge axis, thus we must have a sum of 
\textcolor{black}{
$n_\text{op}$
}
different paths in order to achieve this invariance.

\textcolor{black}{
In Fig.  \ref{fig:operators1} we show the denomination of the operator $O(l_1, l_2)$, the distance $l$ between the straight section of the Wilson line and the charge axis, and the number of terms $n_\text{op}$ in the operator to make it symmetric. 
For instance the first operator $O(0,0)$ is the simple strait Wilson line directly connecting the two charges, it has $l=0$ and just $n_\text{op}=1$ term. 
The next operators are deformations of the strait Wilson line.
Then the operator $O(1,0)$ is a staple with $l=1$, but we need to add $N_\text{op}=4$ of these staples in different directions to have and invariant operator for rotations of angles multiple of $\pi /2$ around the charge axis. The subsequent operator  $O(1,1)$ has its straight $z$ direction line at distance $l=\sqrt 2$ from the charge axis, but for it to be invariant for the two-dimensional rotation around the charge axis and invariant for the parity inversion about the median point we need to have a sum of all $n_\text{op}=8$ of these possible Wilson curves to get a fully symmetric operator. Continuing along this way we can construct an infinite tower of symmetric operators. 
}

Obviously, we must truncate this sum up to some distance $l_\text{max}$ of the order of half the spatial length of the lattice since our lattice has periodic boundary conditions. Ideally, we should have $l_\text{max}$ larger than the finite width of the flux tubes.

%
%--------------
\begin{figure}[t!]% no figure before 1st section
\begin{centering}
\includegraphics[width=0.95\columnwidth]{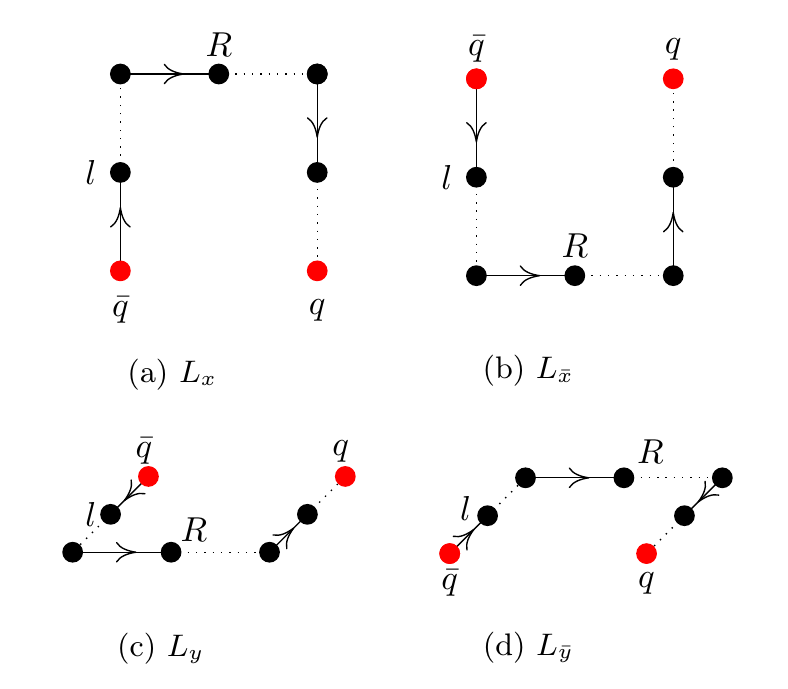}
\caption{For completeness, we show the four sub-operators, spatial Wilson line paths from the antiquark to the quark, used to construct the gauge field operators $O(l,0)= {1 \over \sqrt 4} \left( L_x+L_{\bar x}+L_y+L_{\bar y} \right)$ at Euclidean time $t_0$. The inverse Wilson lines are used for the operators at time $t$.
\label{fig:MO0}}
\par\end{centering}
\end{figure}

However, in contradistinction with the continuum, in a cubic lattice there is a mixing between particular different angular momenta $\Lambda$  operators. Let us consider for instance the operators such as the $O(l_1, l_2)$, which are a linear combination of sub-operators pointing in four directions separated by right or flat angles of $0, \, \pi/ 2, \, \pi, \, 3 \pi /2 $. 
An operator with planar angular momentum $\Lambda$ must have  \cite{Capitani:2018rox}, for each of its spatial links as described for instance in Fig. \ref{fig:operators1}, the phase 
\be  
\textrm{exp}\left( \pm 2 \pi i \,  \Lambda \, \varphi \right) ,
\ee
 where $\varphi$ if the azimuthal angle corresponding to the respective spatial link. In the four directions of our operator, any angular momentum multiple of four,
\be
\Lambda= 0, \, 4, \, 8 \cdots
\ee
has the same phase in all four directions. Thus we must be aware of this problem and do our best to mitigate it. This is further complicated by the phenomenology of flux tubes: in the Nambu-Goto model there is as well a degeneracy between the excited states of these operators.

In particular, we must be careful to prevent the combinations of the symmetric operators $O(l_1, l_2)$ to generate other symmetries. These operators are discrete, they include in general Wilson lines in four to eight directions and all these directions have the same phase. Having the same phase in all directions is necessary for a fully symmetric $\Sigma_g^+$ operator. However, when combining different operators with different directions, a non-symmetric operator may be generated. For instance $O(1,0)$ which azimuthal  directions in the $xy$ plane are $\varphi=0, \, \pi/ 2, \, \pi, \, 3 \pi /2 $ and 
$O(1,1)$ which azimuthal directions in the $xy$ plane are $\varphi=\pi /4, \, 3 \pi/ 4, \, 5 \pi /4, \, 7 \pi /4 $ may be combined   with an opposite phase, for instance $O(1,0) - O(1,1)$ in the case they are both properly normalized, which would correspond to a  $\Gamma_g$ state.

Indeed if these $O(l_1, l_2)$ operators are all included in a correlation matrix, its diagonalization will pick up not only $\Sigma_g^+$ states but also states with large angular momentum $\Lambda=  4$ about the charge axis.
We explicitly verified that, when using a wider basis of operators $O(l_1,l_2)$, we would get more energy levels, some of them nearly degenerate, than expected. Besides, the signal of the excited potentials would be less clear.

We thus restrict our basis of operators to avoid as much as possible states other than $\Sigma_g^+$. We only consider operators which set of Wilson curves have the same directions. It is convenient to use the operators on the directions $x$ and $y$ of the lattice axes. 
For completeness, we show in Fig. \ref{fig:MO0} the four sub-operators, spatial Wilson line paths from the antiquark to the quark, used to construct the gauge field operators 
\be
O(l,0)= {1 \over \sqrt 4} \left( L_x+L_{\bar x}+L_y+L_{\bar y} \right)
\ee
at Euclidean time $t_0$. The inverse Wilson lines are used for the operators at time $t$.

Our basis to construct the correlation matrix consists of the operators of type $O(l,0)$ only, but considering several different $l$. 
%It is interesting that we verified in our lattice computation, we checked such using a smaller basis of operators, less dense in the space of the flux tube, we get clearer results for the spectrum of excited states.
It is interesting that using a smaller operator basis and less dense in the space of the flux tube lead to clearer results for the spectrum of excited states as we found out in our computation.

%SSSSSSSSSSSSSSSSSSSSSSSSSSSSSSSSSSSSSSSSSSSSSSSSSSSSSSSSSSSSSSSSSSSSSSSS
%SSSSSSSSSSSSSSSSSSSSSSSSSSSSSSSSSSSSSSSSSSSSSSSSSSSSSSSSSSSSSSSSSSSSSSSS
%SSSSSSSSSSSSSSSSSSSSSSSSSSSSSSSSSSSSSSSSSSSSSSSSSSSSSSSSSSSSSSSSSSSSSSSS
\subsection{Solving the GEVP for the correlation matrix to compute the excited spectra \label{sec:spectra}}

%--------------
\begin{table*}[t!]
\begin{ruledtabular}
\begin{tabular}{ccc|ccccccccc}
ensemble  & action & operators & $\beta$ & Volume & $u_s$& $u_t$ & $\xi$ & $\xi_R$ & $a_s\sqrt{\sigma}$ & $a_t\sqrt{\sigma}$ & \# configs \\
\hline
$O_1$	& Wilson & 11 $O(l_1,l_2)$ & 6.2 & $24^3\times 48$ & - & - & 1 & 1 & $0.1610$ &  $0.1610$ & 1180 \\
$O_2$	& 	Wilson & 11 $O(l_1,l_2)$ & 5.9 & $24^3\times 48$ & - &  - & 2  & 2.1737(4) & $0.3088(4)$ & $0.1421(2)$ & 2630 \\\hline
%	Wilson no smear & 6.2 & $24^3\times 48$ & - & - & - & - & & & 4000 \\
$W_1$	& 	Wilson & 13 $O(l,0)$ & 6.2 & $24^3\times 48$ & - & - & 1 & 1 & $0.1610$ &  $0.1610$ & 2500 \\
$W_2$	&	Wilson & 13 $O(l,0)$ & 5.9 & $24^3\times 48$ & - &  - & 2  & 2.1737(4) & $0.3093(2)$ & $0.1423(1)$ & 2170 \\
$W_4$	&	Wilson & 13 $O(l,0)$ & 5.6 & $24^3\times 96$ & - & - & 4 & 4.5459(9) & $0.4986(4)$ & $0.1097(1)$ & 3475 \\
$S_4$	&	$S_{II}$ & 13 $O(l,0)$ & 4.0 & $24^3\times 96$ & 0.82006 & 1.0 & 4 & 3.6266(32) & $0.3043(3)$ & $0.0839(1)$ & 3575 \\
\end{tabular}
\end{ruledtabular}
\caption{Our ensembles, for the isotropic Wilson action and the improved anisotropic $S_{II}$ action. $\xi$ is the bare anisotropy in the Lagrangian and $\xi_R$ is the renormalized anisotropy. The renormalized anisotropy and the lattice spacings are computed with the prescription of Section \ref{sec:renormalized}.
}
\label{tab:ensemb}
\end{table*}

\begin{comment}
We utilise the correlation matrix $\langle {\mathcal C}_{kl}(t) \rangle$ to compute the energy levels of the excited states, as done previously in the literature \cite{Blossier:2009kd,Dudek:2009qf,Dudek:2010wm,Bicudo:2021qxj}. 
Now the sub-indices $k$ and $l$ stand for the spacial operators in the operator basis defined in Section  \ref{sec:operator}, denoted $O_k$.
The spacial operators, defined in Figs. \ref{fig:exoperators},  \ref{fig:operators1},  \ref{fig:MO0},  are connected by temporal Wilson lines $L$,
\begin{eqnarray}
 {\mathcal C}_{kl}(t) &=&
 O_k(-\mathbf R/2, \mathbf R/2, 0) \, 
 L(\mathbf R /2, 0, t) \, 
\\
 \non
 &&
 O_l^\dagger(-\mathbf R/2,  \mathbf R /2, t) \,   
 L^\dagger(- \mathbf R /2, 0, t) \ .
\end{eqnarray}
The statistical average $\langle \cdots \rangle$ is performed over our ensemble of gauge link configurations.
\end{comment}

We utilise the correlation matrix $\langle {\mathcal C}_{kl}(t) \rangle$ to compute the energy levels of the excited states, as done previously in the literature \cite{Blossier:2009kd,Dudek:2009qf,Dudek:2010wm,Bicudo:2021qxj}. 
Now the sub-indices $k$ and $l$ stand for the spacial operators in the operator basis defined in Section  \ref{sec:operator}, denoted $O_k$.
The spacial operators, defined in Figs. \ref{fig:exoperators},  \ref{fig:operators1},  \ref{fig:MO0},  are connected by temporal Wilson lines $L$,
\begin{eqnarray}
 {\mathcal C}_{kl}(t) &=&\langle
 O_k(-\mathbf R/2, \mathbf R/2, 0) \, 
 L(\mathbf R /2, 0, t) \, 
\\
 \non
 &&
 O_l^\dagger(-\mathbf R/2,  \mathbf R /2, t) \,   
 L^\dagger(- \mathbf R /2, 0, t)\rangle \ .
\end{eqnarray}
%The statistical average $\langle \cdots \rangle$ is performed over our ensemble of gauge link configurations.

Notice each matrix element corresponds to an evolution operator in Euclidean space, where all energy levels $E_i$ contribute, with coefficients depending on how close the operator is to the actual physical states, with the  Euclidean damping factor $\exp( -E_i \, t)$.

The first step to compute the energy levels, is to diagonalise the Generalized Eigenvalue Problem for the correlation matrix 
\be
{\mathcal C}(t) v_n (t, t_0 ) = \lambda_n (t, t_0 ) {\mathcal C}(t_0 ) v_n (t_0) \ ,
\label{eq:GEVP}
\ee
for each time extent $t$ of the Wilson loop, and get a set of time dependent eigenvalues $\lambda_i(t)$.
With the time dependence, we study the effective mass plot
\begin{equation}
E_i \simeq  \log { \lambda_i(t) \over \lambda_i(t+1)} \ ,
\end{equation}
and search for clear plateaux consistent with a constant energy $E_i$ in intervals $ t \in [{t_i}_\text{ini}, {t_i}_\text{fin}]$ between the initial and final time of the plateau. We have the option of choosing the initial time and we choose $t_0=0$ because it produces the clearest results. Moreover, the results with  $t_0=0$ are compatible with the ones obtained with other small values of $t_0$.

The different energies  levels $E_i$, should correspond to the groundstate and excited states of the flux tube. If our operator basis is good enough, then $E_0$ is extremely close to the groundstate energy, $E_1$ is very close to the the first excited state, etc.

Moreover, with the diagonalization 
we also obtain the eigenvector operators \cite{Bicudo:2021qxj} corresponding to the groundstate, first excitation, etc. 
We get a linear combination of our initial operators,
\begin{eqnarray}
\label{eq:eigen}
\widetilde O_0=c_{01}\, O_1+ c_{02} \, O_2 + \cdots
\\
\nonumber
\widetilde O_1=c_{11} \, O_1+ c_{12} \, O_2 + \cdots
\\
\nonumber
\cdots
\end{eqnarray}
Notice this result must be interpreted with a grain of salt. 
The eigenvector operators $\widetilde O_i$ do not exactly correspond to the respective state as in quantum mechanics,
but they get the clearest possible signal to noise ratio, among our operator basis.

The eigenvector operators $\widetilde O_i$ and the respective correlation matrix can be used in the same time interval $ t \in [{t_i}_\text{ini}, {t_i}_\text{fin}]$ ideal for the effective mass plateaux of the energy spectrum.

The number of gluonic operators  turns out to be large, requiring a large computer power. We thus write all our codes in CUDA and run them in computer servers with NVIDIA GPUs. Due to the GPU limited memory this requires an intensive use of atomic memory operations. 
For instance, to compute a $13\times 13$ correlation matrix, %for a given time $t$ and distance $\mathbf R$,
 per configuration, our CUDA code takes approximately 380s for a lattice volume of $24^3\times 96$ using a GeForce RTX 2080 Ti with 7.5cc.
 %\textcolor{black}{XXX update needed XXX} min to run on a \textcolor{black}{XXX update needed XXX
%$GeForce RTX 2080 Ti, 7.5cc
%O(i) 13 operators
%350-382s
%O(i,j) 11 operators
%171s
%}

%SSSSSSSSSSSSSSSSSSSSSSSSSSSSSSSSSSSSSSSSSSSSSSSSSSSSSSSSSSSSSSSSSSSSSSSS
%SSSSSSSSSSSSSSSSSSSSSSSSSSSSSSSSSSSSSSSSSSSSSSSSSSSSSSSSSSSSSSSSSSSSSSSS
%SSSSSSSSSSSSSSSSSSSSSSSSSSSSSSSSSSSSSSSSSSSSSSSSSSSSSSSSSSSSSSSSSSSSSSSS
\subsection{Gluon actions and configuration ensembles  \label{sec:efficiency}}

We compute our results using six different ensembles, defined in Table \ref{tab:ensemb}.

We use the anisotropic Wilson action \cite{Wilson:1974sk} computed with plaquettes, 
\begin{equation}
	S_\text{Wilson} = \beta\left(\frac{1}{\xi} \sum_{x,s>s'} W_{s,s'} + \xi \sum_{x,s} W_{s,t}  \right)
\end{equation}
where  $W_c = \sum_c {1\over3} \text{Re Tr}(1 - P_c )$  
where $s$, $s'$ runs over spatial links in different positive directions, $P_{s,s'}$ denotes the spatial plaquette, $P_{s,t}$ the spatial-temporal plaquettes and $\xi$ is the (unrenormalized) anisotropy.

Moreover, to improve our signal we also resort to the improved anisotropic action $S_{II}$ developed in Ref. \cite{Morningstar:1996ze}, 
\bea
S_\text{II} &=& \beta \left( \frac{1}{\xi}     \sum_{x,s>s'}  \left[\frac{5 W_{s,s'}}{3 u_s^4} - \frac{W_{ss,s'}+W_{s's',s}}{12 u_s^6} \right]+   \right.
\non \\
& &\left. + \xi \sum_{x,s} \left[\frac{4 W_{s,t}}{3 u_s^2 u_t^2} - \frac{W_{ss,t}}{12 u_s^4 u_t^2} \right]   \right),
\eea
with $u_s = \langle {1 \over 3} \text{Re Tr} P_{ss} \rangle^{1/4}, \ u_t=1$.
$W_{ss,s'}$ and $W_{ss,t}$, instead of plaquettes, include $2\times 1$ rectangles.

So far, the results with more excited states shown in the literature, two states 
\cite{Juge:1999ie,Juge:2002br} and up to three in an unpublished work 
\cite{Morningstar:website}, have been obtained with this action.

We generate five different ensembles of configurations, with the parameters defined of Table \ref{tab:ensemb}.

The  anisotropy is used in order to have a smaller temporal lattice spacing $a_t$. 
This enables a better estimation on the extraction of excited states as well as a more precise result since we have more time slices for the same time intervals.
%It is more adequate, not only to the study of excited sates, but also for more precise results since we have more time slices for the same time intervals. 

Notice an anisotropic action enables us to use larger distances with the same number of spatial lattice points. However, in order to have a good spatial resolution, it is then convenient to use an improved action, and  $S_{II}$ uses plaquettes extended up to $1 \times 2$ rectangular shapes. $S_{II}$ is specially designed to eliminate spurious high-energy states from the gluon spectrum.

Moreover, in order to improve the signal over noise ratio, for the Wilson ensemble, we use the multihit technique in the temporal Wilson lines and the APE smearing in the spatial Wilson lines \cite{Cardoso:2013lla}.
The multihit technique,  \cite{Brower:1981vt, Parisi:1983hm}, replaces each temporal link by its thermal average,
\begin{equation}
	U_4\rightarrow \bar{U}_4=\frac{\int dU_4 U_4 \,e^{\beta\Tr \left[U_4 F^\dagger\right]}}{\int dU_4 \,e^{\beta\Tr\left[ U_4 F^\dagger\right]}} \ .
\end{equation}
Here it is not possible to utilise the extended Multihit technique as defined in Ref.  \cite{Cardoso:2013lla}, because our operators in the spatial Wilson line have a broader structure.
In particular, we use MultiHit with 100 iterations in time followed by APE smearing \cite{Albanese:1987ds} with $\alpha=0.4$ and 20 iterations for the Wilson action without anisotropy.
For the $S_{II}$ ensemble and for the Wilson ensembles with anisotropy, we use MultiHit with 100 iterations in time followed by Stout smearing \cite{Morningstar:2003gk} in space with $\alpha=0.15$ and 20 iterations.

It turns out it is more economical, using GPUs, to perform all our computations on the fly, rather than saving configurations. In general, we first generate a configuration, then apply smearing, then compute the correlation matrix with our full operator basis. What we save to disk is the correlation matrix. Since we generate the computations on the fly, we also list  the ensemble Table \ref{tab:ensemb} the sets of operators used. Notice we may as well turn off smearing, to check the importance of smearing, and we study the results of ensembles $W_1$, $W_2$, $W_4$ and $S_4$ with and without smearing.

In what concerns the efficiency of our codes, for instance using a lattice volume of $24^3\times 96$ and the Wilson action, it takes 
5s to generate a new configuration. To decorrelate the configurations, we run 50 iterations between   the used configurations. Each iteration is composed by 4 heatbath steps followed by 7 overrelaxation steps. 
It takes 0.8s for 100 iterations of Multihit and 
0.014s for 20 steps of Stout smearing in space.

%SSSSSSSSSSSSSSSSSSSSSSSSSSSSSSSSSSSSSSSSSSSSSSSSSSSSSSSSSSSSSSSSSSSSSSSS
%SSSSSSSSSSSSSSSSSSSSSSSSSSSSSSSSSSSSSSSSSSSSSSSSSSSSSSSSSSSSSSSSSSSSSSSS
%SSSSSSSSSSSSSSSSSSSSSSSSSSSSSSSSSSSSSSSSSSSSSSSSSSSSSSSSSSSSSSSSSSSSSSSS
\subsection{Computing the lattice spacing and the renormalized anisotropy  \label{sec:renormalized}}

In the case of isotropic actions, there is only one independent scale, arising from dimensional transmutation, since the action is conformal invariant. The physical scale is the string tension $\sigma$ present in the linear term of the quark-antiquark potential. From its value, we determine the lattice spacing $a$.
In the case of anisotropic actions we have two different lattice spacings, the spatial $a_s$ and the temporal $a_t$.

To set the scale of the lattice spacing $a$ of the isotropic Wilson action (with $\xi=1$) in physical units, corresponding to the ensemble $W_1$, we use the equations fitted in SU(3) pure gauge lattice QCD by Ref. \cite{Edwards:1997xf},
\begin{eqnarray}
\left(a\sqrt\sigma\right)(g) & = & f(g^2) \left[ 1 + b_1 \,\hat{a}(g)^2 + b_2 \,\hat{a}(g)^4 +\right.\nonumber \\
 &  & \left. + b_3 \,\hat{a}(g)^6 \right]/b_0 ,
\end{eqnarray}
where $g$ is the coupling constant of the Wilson action,
and  \cite{Edwards:1997xf}  the parameters are $b_0 = 0.01364$, $b_1 = 0.2731$, $b_2 = -0.01545$ and $b_3 = 0.01975$, valid in the region $5.6 \leq \beta \leq 6.5$,
\begin{equation}
\hat{a}(g) = \frac{f(g^2)}{f(g^2(\beta=6.0))},
\end{equation}
\begin{equation}
f(g^2) = \left(b_0g^2\right)^{-\frac{b_1}{2b_0^2}} \exp\left( -\frac{1}{2b_0g^2}  \right),
\end{equation}
and
\begin{equation}
b_0 = \frac{11}{(4\pi)^2},\quad\quad b_1 = \frac{102}{(4\pi)^4} \ .
\end{equation}

As for anisotropic actions, the renormalized anisotropy $\xi_R=a_s/a_t $ can be determined as a function of the bare anisotropy $\xi$. The groundstate potential is computed with the Wilson loop, two different directions for the time direction, either the usual direction 4 now anisotropic, or using one of the isotropic distances, say 3. Comparing the short distance potential with both time directions, which is well determined without any smearing and is fitted with a simple function, the ratio $a_s/a_t$ is determined.

For the pure gauge SU(3) anisotropic Wilson action, ensembles $W_2$ and $W_4$, Ref. \cite{Drummond:2002yg, Drummond:2003qu} fitted the parametrisation,
\be
	\xi_R = \left(1+\eta(\xi)g^2\right) \xi \ ,
\ee
corresponding to the series,
\be
	Z(g^2,\xi)=\xi_R/\xi=1+\eta(\xi)g^2+\mathcal{O}(g^4) \ .
\ee
where,
\bea
	g^2 &=& \frac{6}{\beta}
\\ \non
	\eta(\xi) &=& 0.1687(2) - 0.16397(4)/\xi - 0.005245(2)/\xi^2 \ .
\eea

For the $S_{II}$ action, ensemble $S_4$, we have similar equations from Ref. \cite{Drummond:2002yg, Drummond:2003qu}, the only difference is in the parameters of,
\be
\eta(\xi_0) = 0.0955(4) - 0.0702(16)/\xi_0 - 0.0399(14)/\xi_0^2,
%	\eta(\xi_0) = 0.0602(1) - 0.0656(2)/\xi_0 - 0.0237(1)/\xi_0^2,
\ee
with $\xi_0=\xi \,u_s/u_t$ and $\xi_R=\left(1+\eta(\xi_0)g^2\right)\xi_0$. 
We explicitly checked these formulas comparing them with the values used in Ref. \cite{Morningstar:website1}. In general both values for $\xi_R$  are compatible within error bars, only in one case the difference is slightly larger than the error bar.

Once the physical anisotropy ratio $\xi_R$ is determined, we determine the physical scale from the string tension. We fit the data for the groundstate potential at large interquark distance $R$ with the linear plus constant ansatz $c_0 + c_1\,R$. 
We did not find any improvement in the results fitting the data with the ansatz $c_0 + c_1\,r + c_2 /r$.
%We did not find it would improve the results to fit the data with the ansatz $c_0 + c_1\,r + c_2 /r$.
We utilise smearing to be able to fit the large distance potential.
%The linear term is obtained from fitting the difference for different distances $R$ the groundstate potential,
The linear term is obtained from fitting the difference for different distances $R$ of the ground-state potential,
\be
c_1=E_0(R+1)-E_0(R)
\ee
with a plateau.
The linear term depends on the spatial lattice spacing $a_s$, the temporal lattice spacing $a_t$ and the string tension $\sigma$ as,
\be
c_1 = a_s a_t \sigma \ , 
\ee
therefore
\begin{eqnarray}
a_s \sqrt{\sigma} & = & \sqrt{c_1 \xi_R}, \\
a_t \sqrt{\sigma} & = & \sqrt{c_1 / \xi_R}. 
\end{eqnarray}
Note that, using the above formulas, all the results presented in this work are in units of $\sqrt{\sigma}$ for the potentials and in units of  $1/\sqrt{\sigma}$ for the distances. This is the standard two step technique to determine the scales of our lattice.
%This is the standard two step technique to determine the scales or our lattice

Besides, there is another option to fit both lattice spacings $a_s$ and $a_t$ in one step, when we fit the whole excited flux tube spectrum. This will be discussed in Section \ref{sec:analysis} when we will analyse the spectrum.

%SSSSSSSSSSSSSSSSSSSSSSSSSSSSSSSSSSSSSSSSSSSSSSSSSSSSSSSSSSSSSSSSSSSSSSSS
%SSSSSSSSSSSSSSSSSSSSSSSSSSSSSSSSSSSSSSSSSSSSSSSSSSSSSSSSSSSSSSSSSSSSSSSS
%SSSSSSSSSSSSSSSSSSSSSSSSSSSSSSSSSSSSSSSSSSSSSSSSSSSSSSSSSSSSSSSSSSSSSSSS
\section{Improved results for the $\Sigma_g^+$ spectrum \label{sec:results}}

Extracting very excited states requires using different state of the art techniques. 
We first show the results of the different techniques we check to improve the signal over noise ratio. Then we show our best results.

%SSSSSSSSSSSSSSSSSSSSSSSSSSSSSSSSSSSSSSSSSSSSSSSSSSSSSSSSSSSSSSSSSSSSSSSS
%SSSSSSSSSSSSSSSSSSSSSSSSSSSSSSSSSSSSSSSSSSSSSSSSSSSSSSSSSSSSSSSSSSSSSSSS
%SSSSSSSSSSSSSSSSSSSSSSSSSSSSSSSSSSSSSSSSSSSSSSSSSSSSSSSSSSSSSSSSSSSSSSSS
\subsection{Comparing the use of several $O(l,0)$ operators with and without smearing of the gauge links \label{sec:opera}}

\begin{figure}[t!]
\begin{centering}
\includegraphics[width=\columnwidth]{{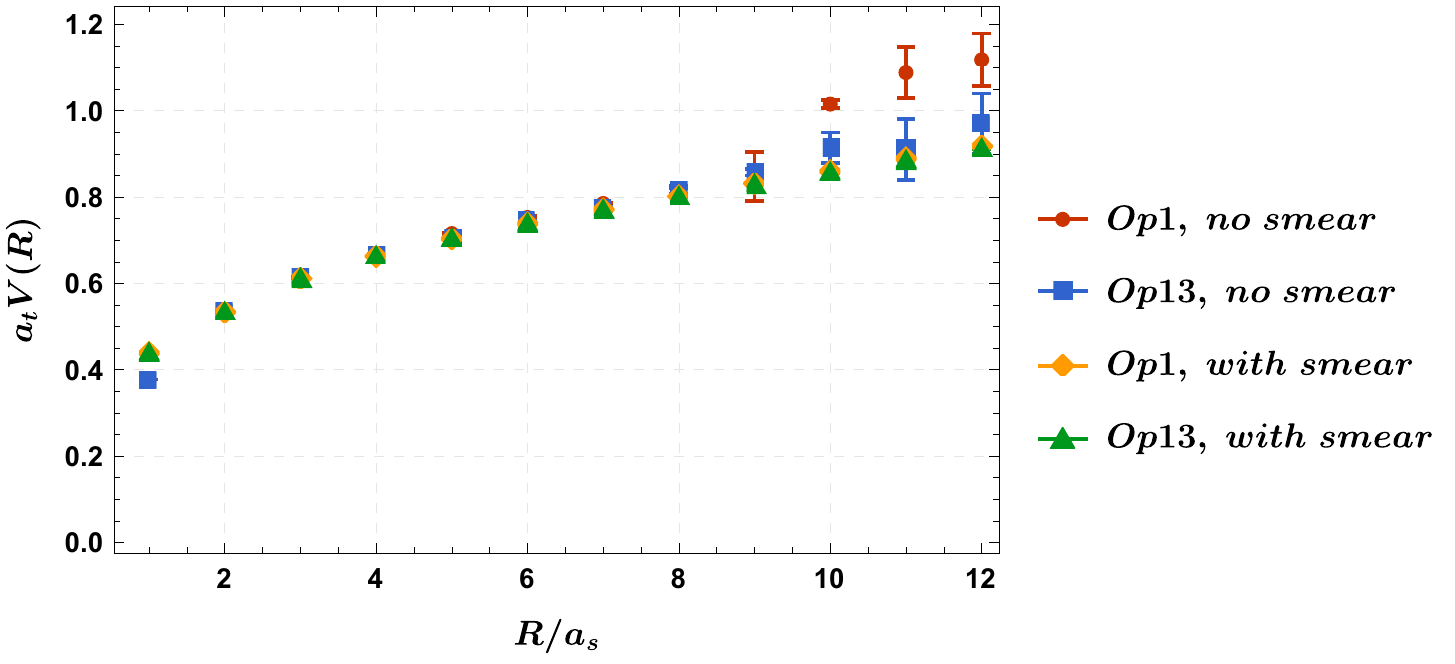}}
\includegraphics[width=\columnwidth]{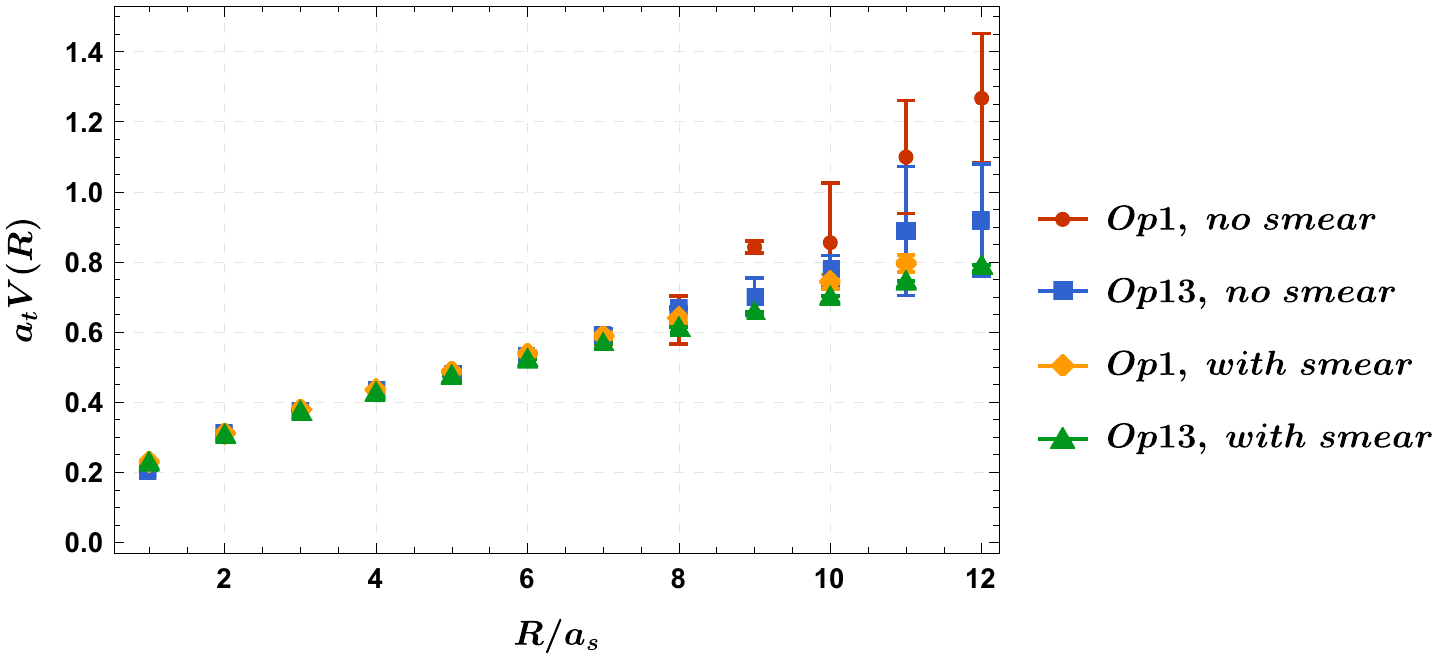}
\includegraphics[width=\columnwidth]{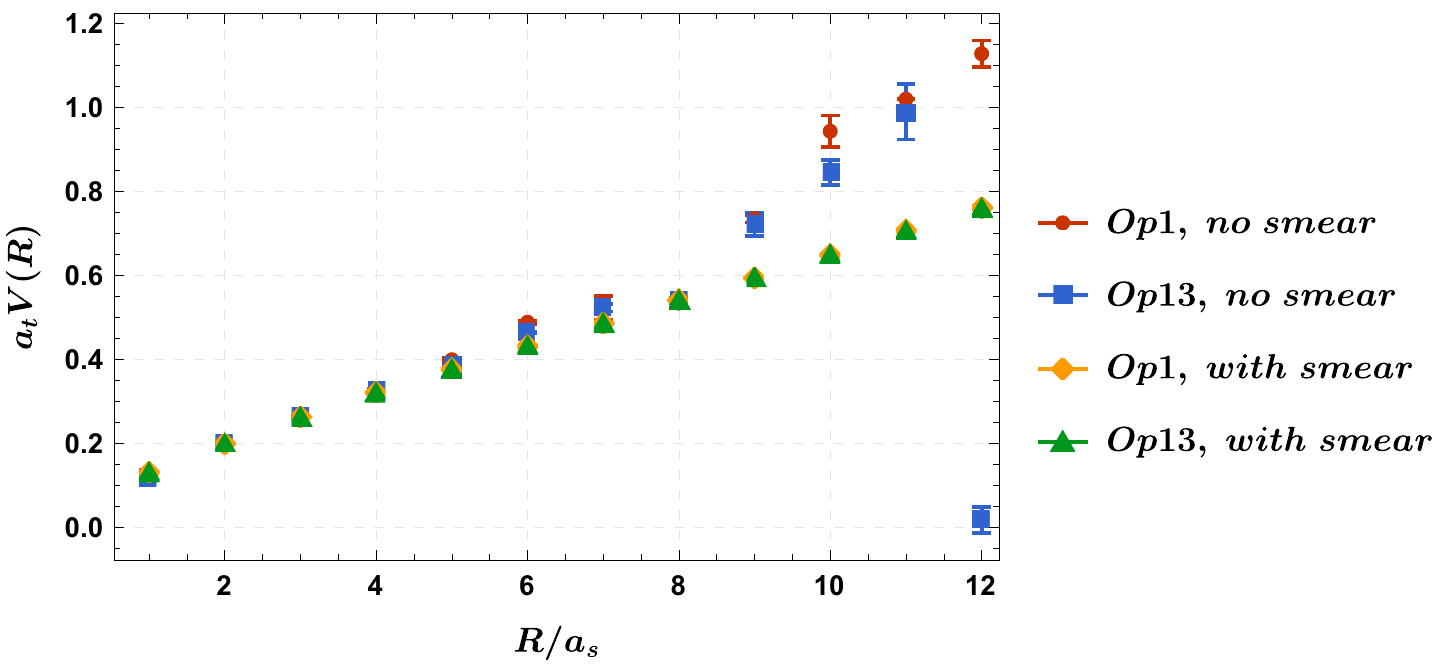}
\includegraphics[width=\columnwidth]{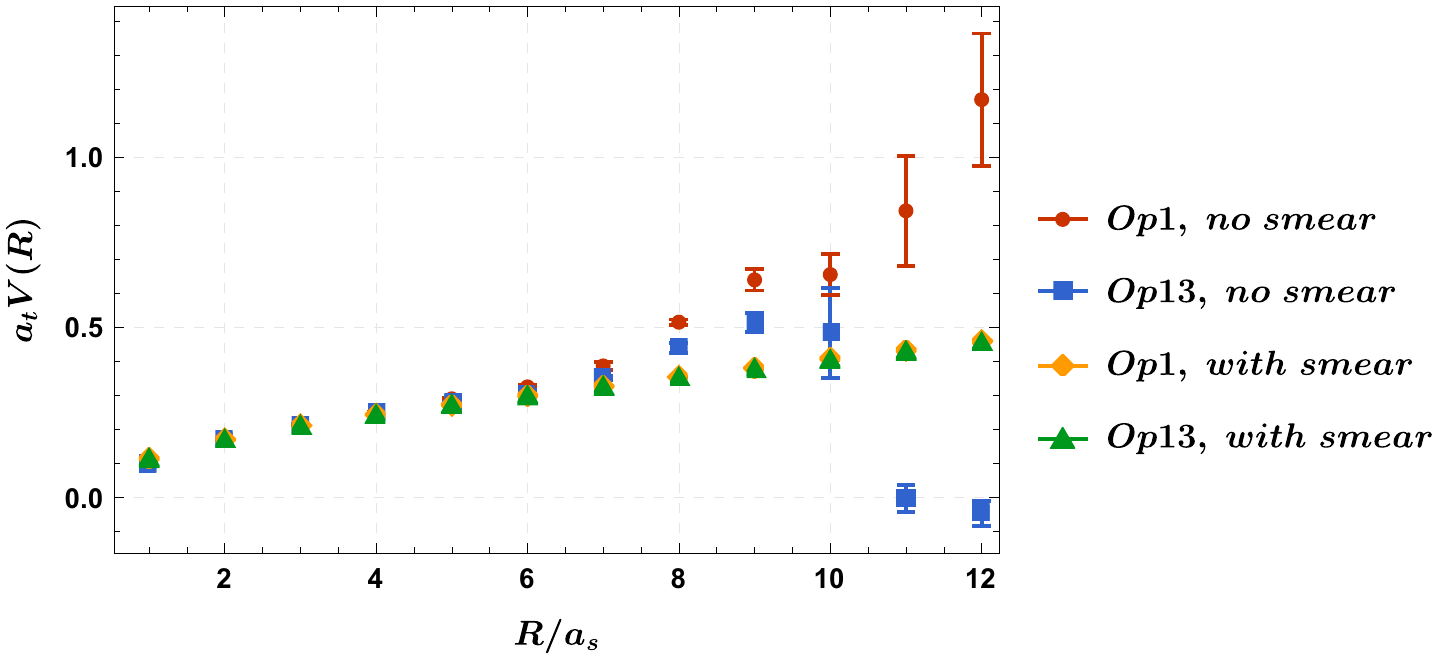}
\par\end{centering}
\caption{Comparing results with and without a basis of operators, with and without smearing. 
Results for the groundstate only, from top to bottom respectively with ensembles $W_1$, $W_2$, $W_4$ and $S_4$.}
\label{fig:pot_nosmear}
\end{figure}

\begin{figure}[t!]
\begin{centering}
\includegraphics[width=\columnwidth]{{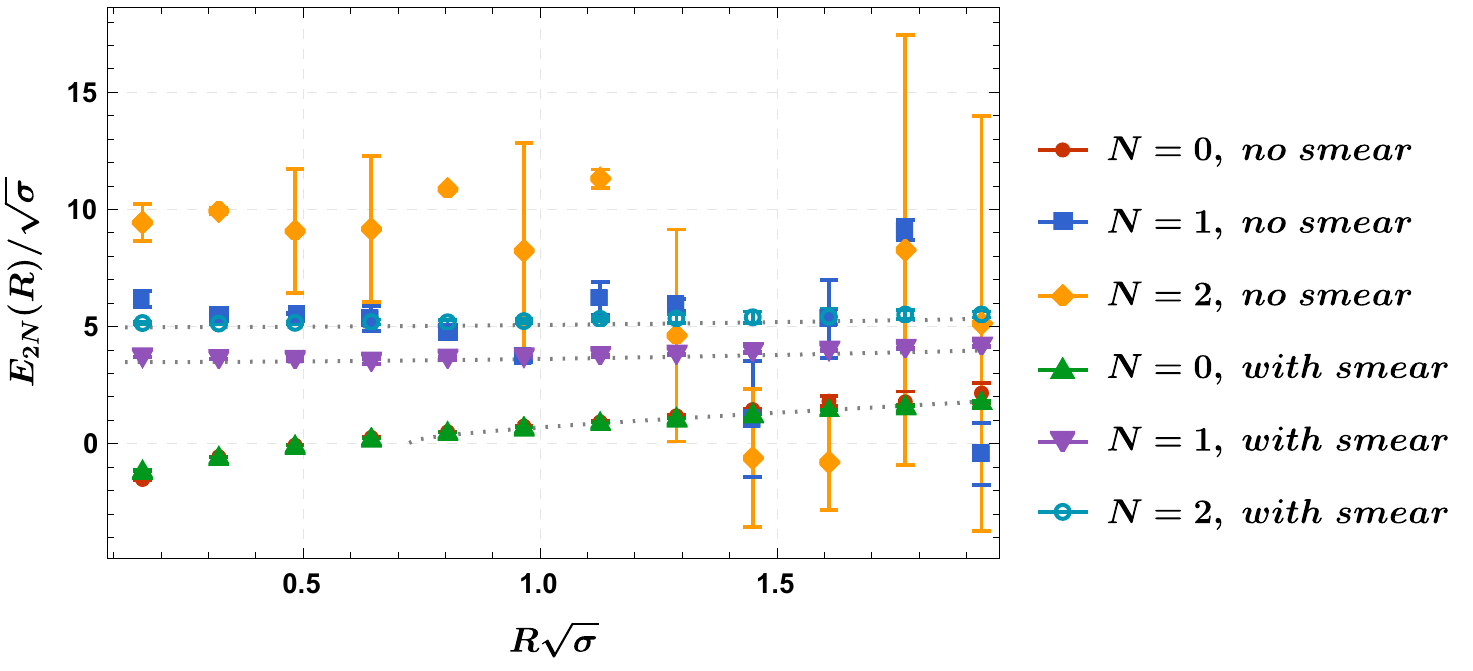}}
\includegraphics[width=\columnwidth]{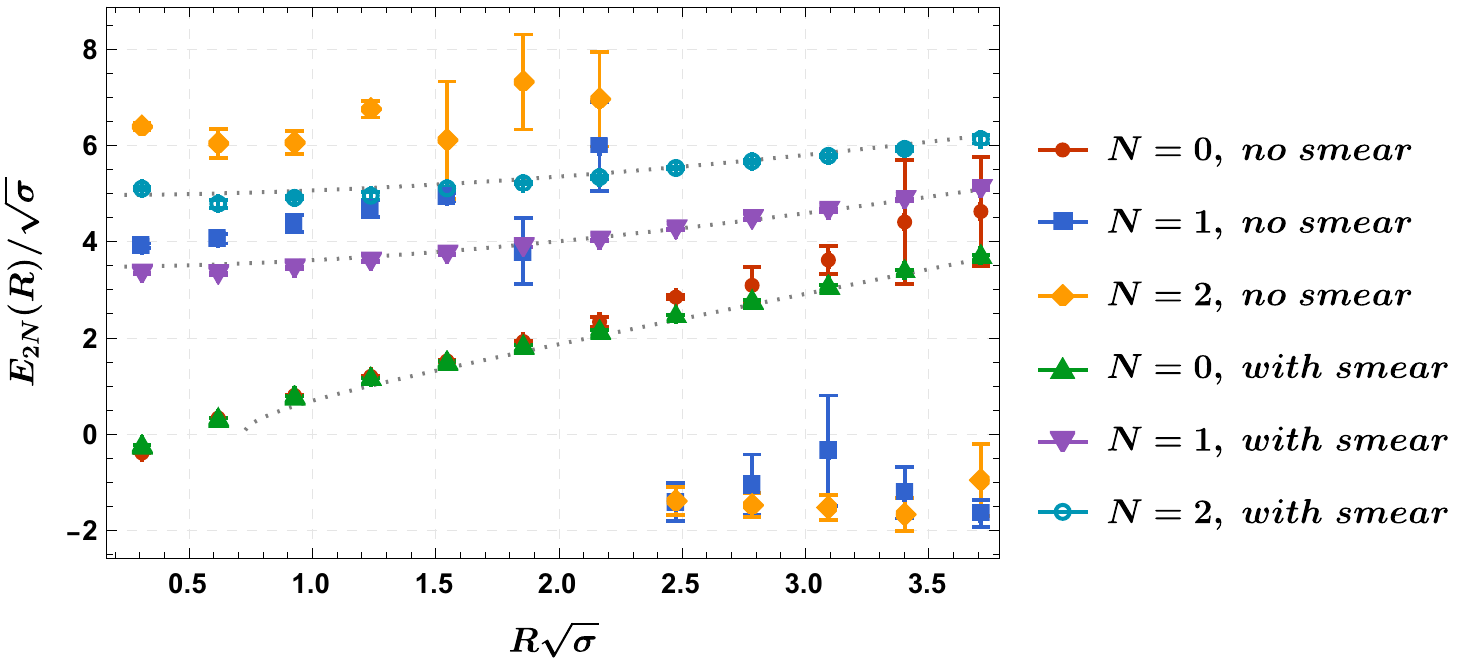}
\includegraphics[width=\columnwidth]{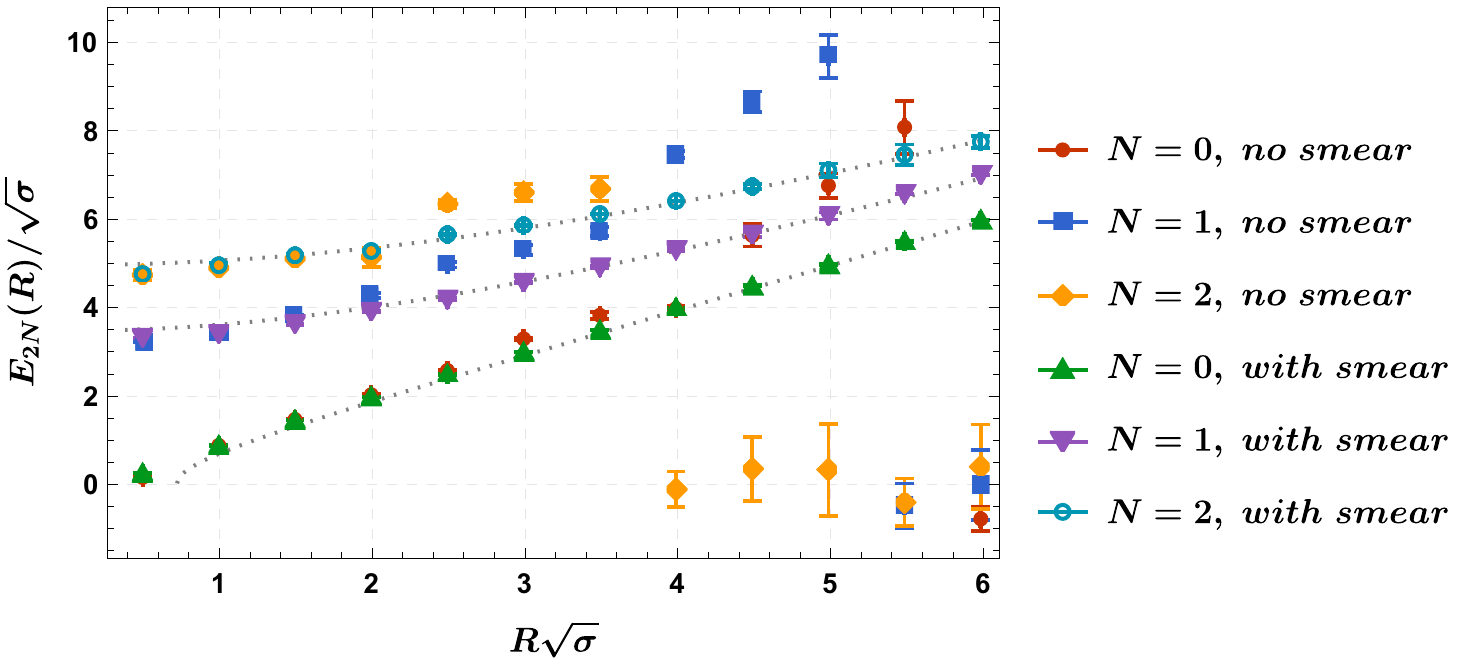}
\includegraphics[width=\columnwidth]{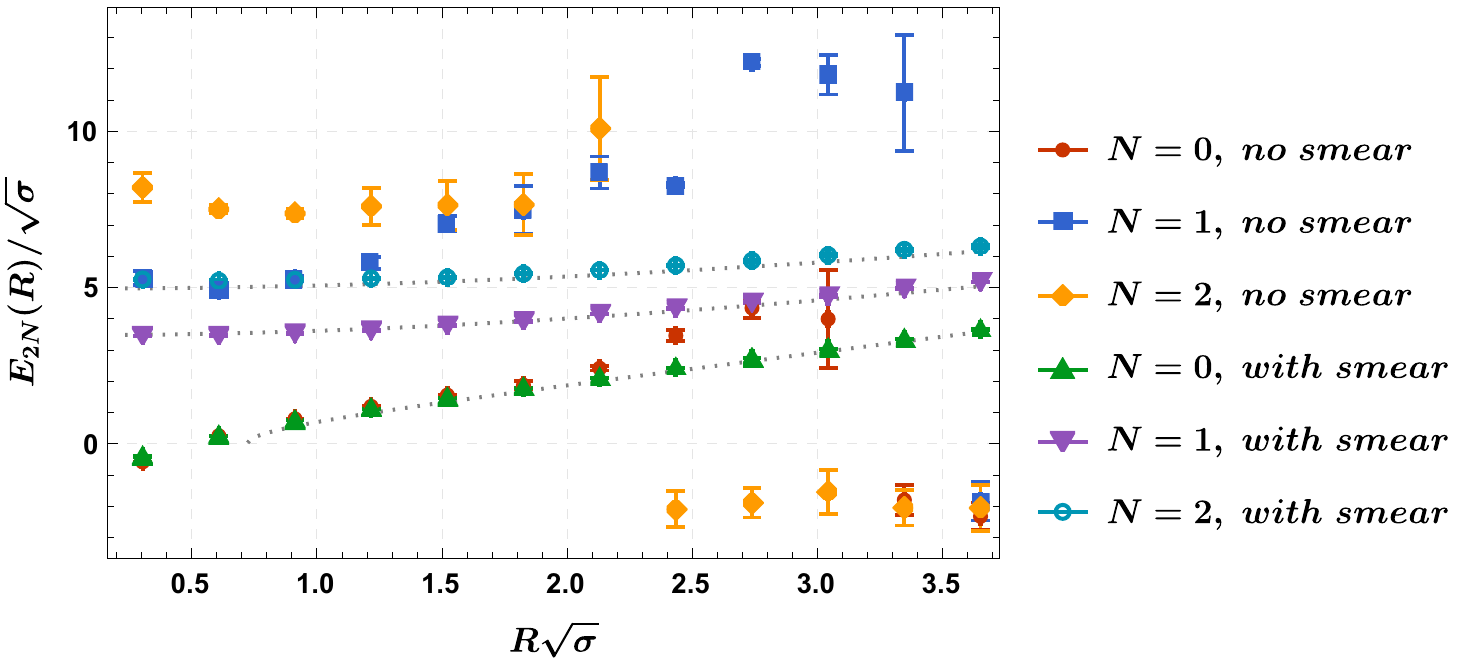}
\par\end{centering}
\caption{
Comparing results with and without smearing, using the 13 operators $O(l,0)$. To guide the eye, the Nambu-Goto model spectrum is shown in dashed lines. Results for the groundstate and first two states, from top to bottom respectively with ensembles $W_1$, $W_2$, $W_4$ and $S_4$.}
\label{fig:pot_nosmear_ex}
\end{figure}

Here we are interested in checking the effect of using a large basis of operators. If this improves the results, it can be used together with other techniques. Notice in general in the literature smearing is mandatory to improve the results, already for the groundstate quark-antiquark fluxtube. We thus compare the effect of using a large basis of operators to using smearing. 

In Fig. \ref{fig:pot_nosmear} we compare the groundstate potential for different ensembles, using just one operator (the standard Wilson loop with $O(0,0)$), or 13 sets of operators of the type $O(l,0)$ up to $l=12$. It turns out the use of 13 operators improves the groundstate signal. But to have the best signal for a distance large enough to be physically interesting, we need smearing. According to the plots, smearing maintains the results unchanged for $R > 1 a_s$, in particular the string tension error bar is reduced with smearing but it remains consistent with the string tension with no smearing. 

In Fig. \ref{fig:pot_nosmear_ex} we use the 13 operators to search for at least one excited state, and we compare the results without and with smearing. We also include the Nambu-Goto spectrum in dashed lines to guide the eye. Indeed the operators are necessary to get the excited sates, but in most cases smearing is necessary as well and produce similar results to Nambu-Goto. Only in ensemble $W_4$, the results for the smaller distances $R < 4 a_s$ can be obtained without any smearing for the first two excited states. Having an excited state with no smearing, similar to the smearing ones and to Nambu-Goto confirms that smearing is not distorting our spectrum. This $W_4$ result also shows that an anisotropic action may be more effective to study excited states.

%SSSSSSSSSSSSSSSSSSSSSSSSSSSSSSSSSSSSSSSSSSSSSSSSSSSSSSSSSSSSSSSSSSSSSSSS
%SSSSSSSSSSSSSSSSSSSSSSSSSSSSSSSSSSSSSSSSSSSSSSSSSSSSSSSSSSSSSSSSSSSSSSSS
%SSSSSSSSSSSSSSSSSSSSSSSSSSSSSSSSSSSSSSSSSSSSSSSSSSSSSSSSSSSSSSSSSSSSSSSS
\subsection{ Degeneracies with  $O(l_1,l_2)$ operators \label{sec:degenera}}

\begin{figure}[t!]
\begin{centering}
\includegraphics[width=\columnwidth]{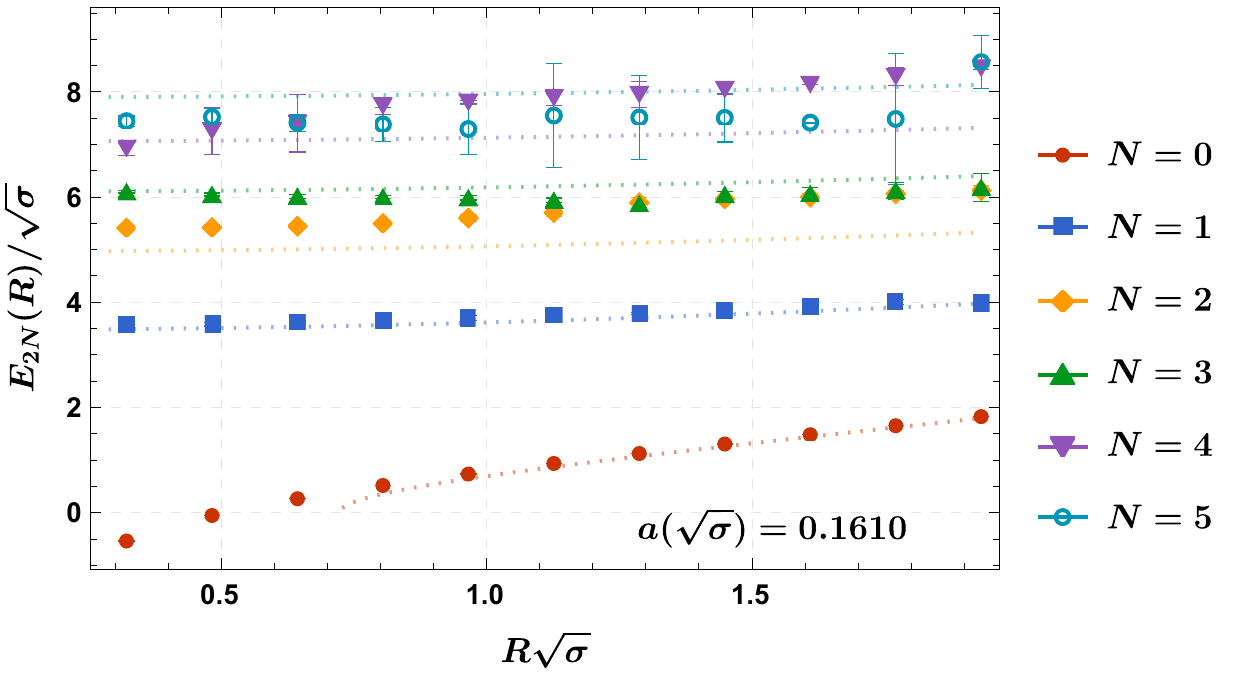}
\includegraphics[width=\columnwidth]{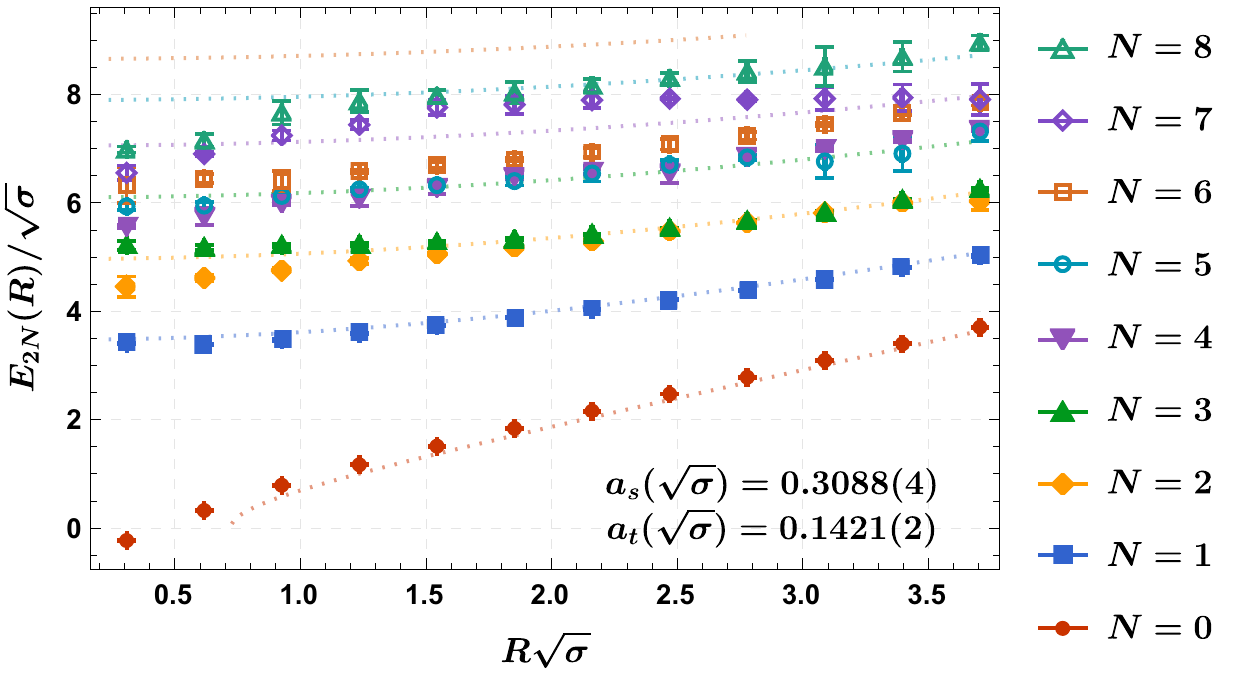}
\par\end{centering}
\caption{Results with the enlarged basis of 11 $O(l_1,l_2)$ operators, for the Wilson gauge action with  APE smearing in space and Multihit in time. To guide the eye, the Nambu-Goto model spectrum is shown in dashed lines. From top to bottom respectively with ensembles $O_1$ and $O_2$. }
\label{fig:pot_operaOl1l2}
\end{figure}

We now study the results obtained with a more complete set of operators, 
using the ensembles $O_1$ and $O_2$ of Table \ref{tab:ensemb}. In these simulations we include  all 11 possible operators $O(l_1,l_2)$  with $l< 4.5$, corresponding to the operators with 
%$l_1\, l_2=0\, 0,1\, 0,2\,0,3\, 0,4\,0,1\, 1,2\, 1,3\, 1,2\, 2,3\, 2,3\, 3$.  
$l_1\, l_2=0, 0;1, 0;2,0;3, 0;4,0;1, 1;2, 1;3, 1;2, 2;3, 2;3, 3$. 
As discussed in Subsection \ref{sec:operator} this may produce states with different symmetries from $\Sigma_g^+$. For instance a phase difference between operators $O(l,0)$ and operators $O(l_1,l_2)$ may be selected by the GEVP, and this could produce operators with angular momentum four, of symmetry $\Gamma_g^+$. 

This is verified in Fig. \ref{fig:pot_operaOl1l2} where  we find more states than in Fig. \ref{fig:pot_opera_1244}, with just the sets of operators $O(l,0)$ which are less prone to generate unwanted symmetries other than  $\Sigma_g^+$. Indeed we find approximately degeneracies among the $N=2, \, N=3 $ states, the $N=4, \, N=5,  \, N=6 $ and the $N=7, \, N=8 $ states.
This makes sense in the Nambu-Goto perspective where there is a principal quantum $n=2*n_r+l$, similar to the principal quantum number for an harmonic oscillator vibrating in a two-dimensional $x-y$ space.

From this perspective we expect to have the degeneracies,
\\
$n=0$, 1 state : $n_r=0$,
\\
$n=2$, 1 state : $n_r=1$,
\\
$n=4$, 3 states: $n_r=2$ or $l=4$, 
\\
$n=6$, 3 states: $n_r=3$ or $n_r=1$ and $l=4$,
\\
$n=8$, 5 states: $n_r=4$ or $n_r=2$ and $l=4$ or $l=8$,
\\
\dots

Indeed we find in Fig. \ref{fig:pot_operaOl1l2} nearly degenerate states $N=2$ and $N=3$, compatible with $n_r=2, \, l=0$ and $n_r=0, \, l=4$ states. They only differ for small distances, where possibly there are Coulomb contributions beyond the Nambu-Goto model. We also find approximate degeneracies for the higher states. 

This confirms as expected that to get clearer and unambiguous $\Sigma_g^+$ signals it is preferable to have a smaller class of operators,  using only the $O(l,0)$ sets of operators.

%SSSSSSSSSSSSSSSSSSSSSSSSSSSSSSSSSSSSSSSSSSSSSSSSSSSSSSSSSSSSSSSSSSSSSSSS
%SSSSSSSSSSSSSSSSSSSSSSSSSSSSSSSSSSSSSSSSSSSSSSSSSSSSSSSSSSSSSSSSSSSSSSSS
%SSSSSSSSSSSSSSSSSSSSSSSSSSSSSSSSSSSSSSSSSSSSSSSSSSSSSSSSSSSSSSSSSSSSSSSS
\subsection{Results with a basis of $O(l,0)$ operators, the Wilson action, and smearing  \label{sec:wilson}}

Using 13 sets of on-axis operators $O(l,0)$ and the Wilson action, with no anisotropy, we find  already several energy levels, clearly ordered. 

\begin{figure}[t!]
\begin{centering}
\includegraphics[width=\columnwidth]{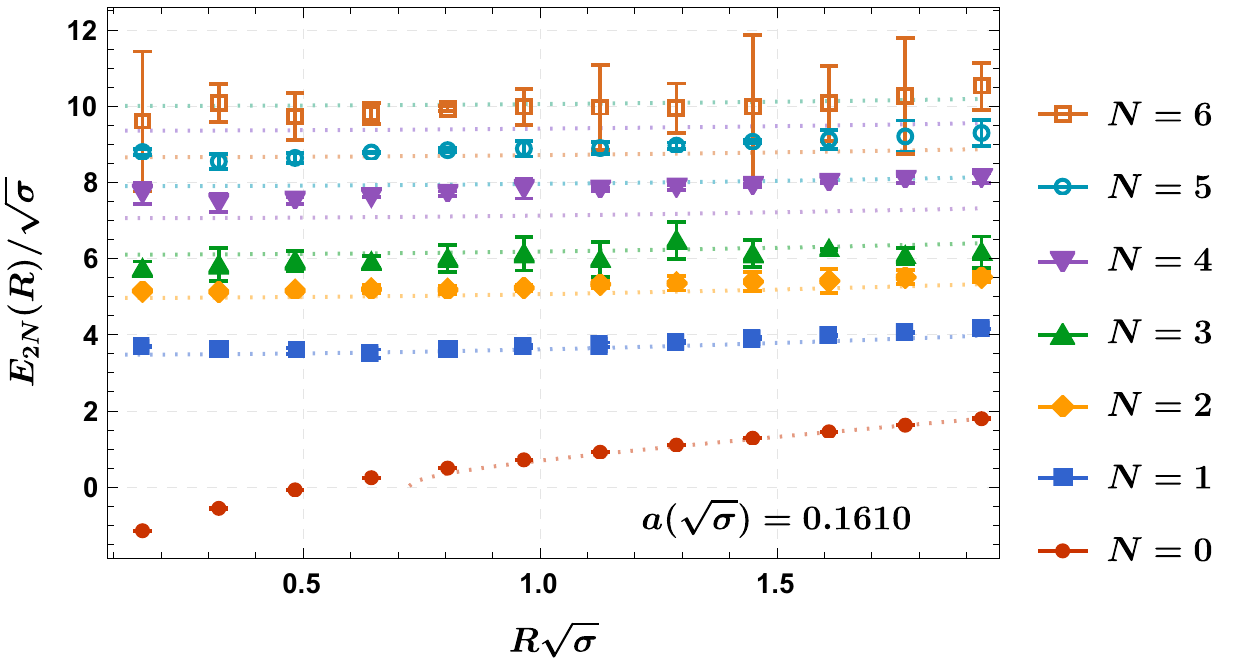}
\includegraphics[width=\columnwidth]{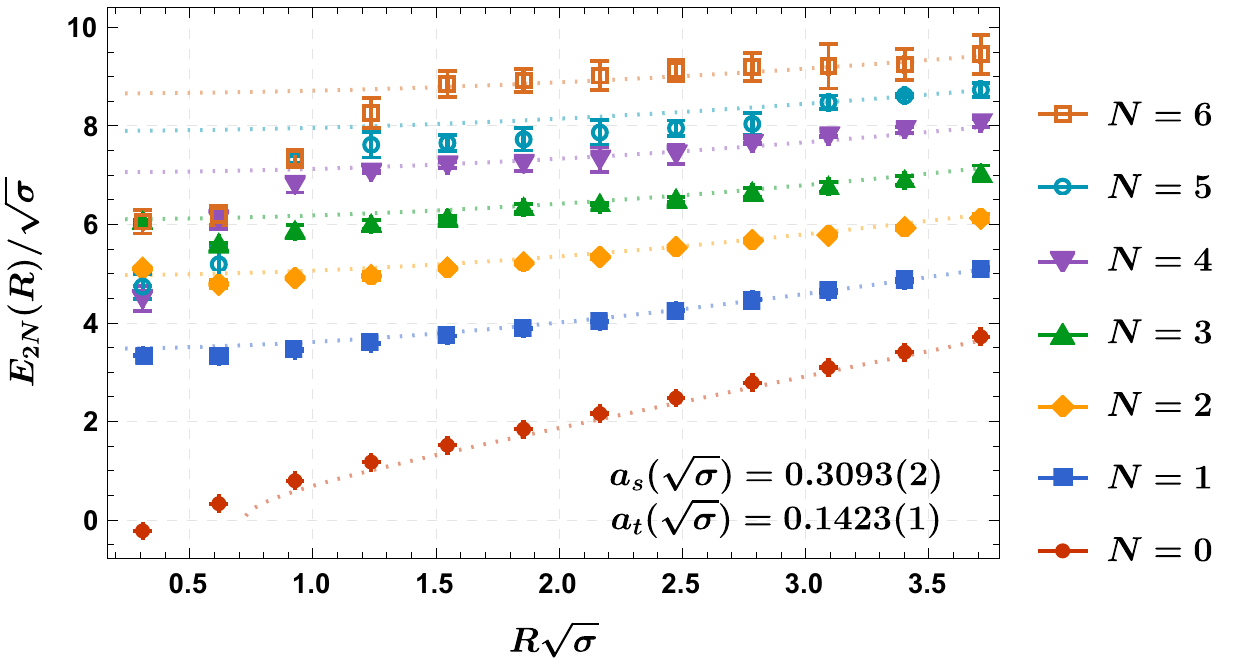}
\includegraphics[width=\columnwidth]{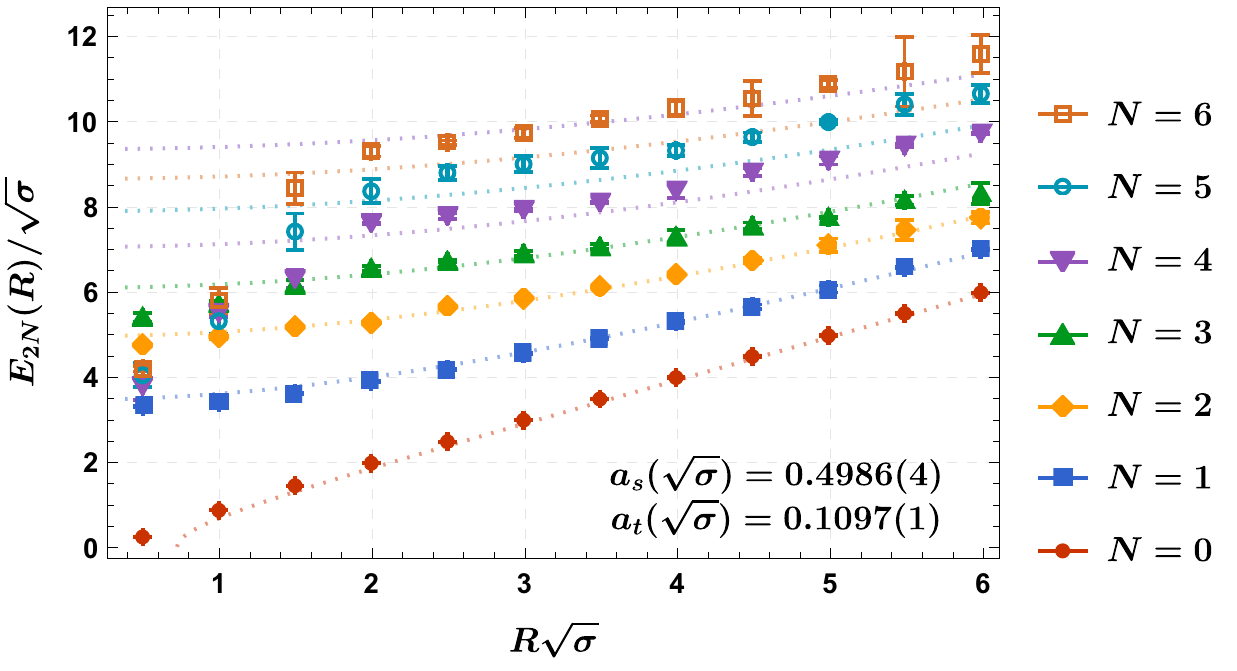}
\includegraphics[width=\columnwidth]{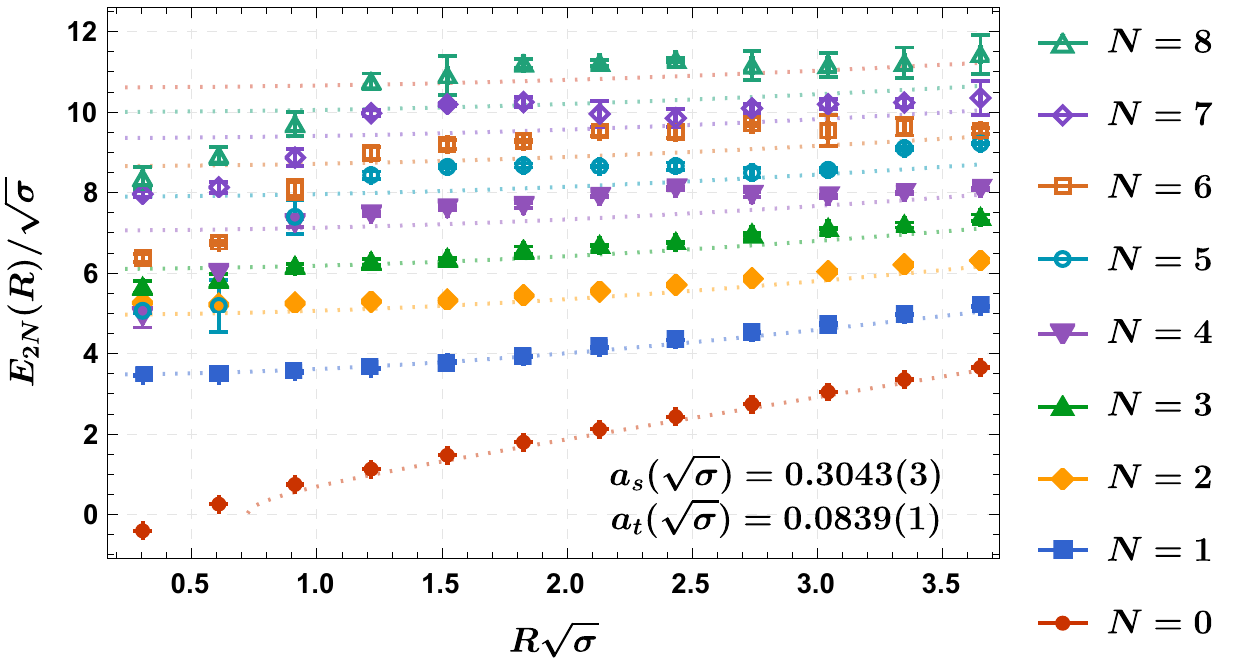}
\par\end{centering}
\caption{Results for the full spectrum, showing only the levels obtained with good Effective Mass Plots, from top to bottom respectively with ensembles $W_1$, $W_2$, $W_4$ and $S_4$. To guide the eye, the Nambu-Goto model spectrum is shown in dashed lines.}
\label{fig:pot_opera_1244}
\end{figure}

However, the level $N=3$ is a bit out of the remaining patterns, with larger error bars a bit unexpectedly closer to the $N=2$ level. This is clear in  the top of Fig. \ref{fig:pot_opera_1244}, ensemble $W_1$.

The energy levels also have a similar dependence on $R$ as in the Nambu-Goto model, constant at short distances and linear at large distances. 

However, the higher levels seem to be shifted vertically when compared with the Nambu-Goto levels depicted with dotted lines.  

%SSSSSSSSSSSSSSSSSSSSSSSSSSSSSSSSSSSSSSSSSSSSSSSSSSSSSSSSSSSSSSSSSSSSSSSS
%SSSSSSSSSSSSSSSSSSSSSSSSSSSSSSSSSSSSSSSSSSSSSSSSSSSSSSSSSSSSSSSSSSSSSSSS
%SSSSSSSSSSSSSSSSSSSSSSSSSSSSSSSSSSSSSSSSSSSSSSSSSSSSSSSSSSSSSSSSSSSSSSSS
\subsection{Results with the Wilson action and anisotropic lattice \label{sec:wilsonaniso}}

To get potentials at large distances, we first use the anisotropic Wilson action. The results are shown in Fig. 
\ref{fig:pot_opera_1244} centre, both for $\xi=2$ and $\xi=4$. Then we are able to go not only up to large distances but also to clearly see one more level, going up to $N=6$.

The pattern of the separation of the levels is also more striking even than in the isotropic case.
%The pattern of the separation of the levels is also more even than in the isotropic case.

However, the values of the higher potentials get distorted at shorter distances, perhaps because the anisotropy somehow enables the degeneracy already discussed in Sections \ref{sec:operator} and \ref{sec:degenera} to set in.

Nevertheless, the higher levels remain undistorted for $R> 4 a_s$.

%SSSSSSSSSSSSSSSSSSSSSSSSSSSSSSSSSSSSSSSSSSSSSSSSSSSSSSSSSSSSSSSSSSSSSSSS
%SSSSSSSSSSSSSSSSSSSSSSSSSSSSSSSSSSSSSSSSSSSSSSSSSSSSSSSSSSSSSSSSSSSSSSSS
%SSSSSSSSSSSSSSSSSSSSSSSSSSSSSSSSSSSSSSSSSSSSSSSSSSSSSSSSSSSSSSSSSSSSSSSS
\subsection{Results with the improved anisotropic action \label{sec:improved}}

To solve the problem at smaller distances for the anisotropic lattices, we resort to the improved action.
Indeed the short distance is improved although it remains distorted, as seen in Fig. \ref{fig:pot_opera_1244} bottom, ensemble $S_4$.

Nevertheless we are able to get two more levels, $N=7$ and 8. This is the highest level we are able to get.

\begin{figure*}[t!]
\begin{centering}
\includegraphics[width=\columnwidth]{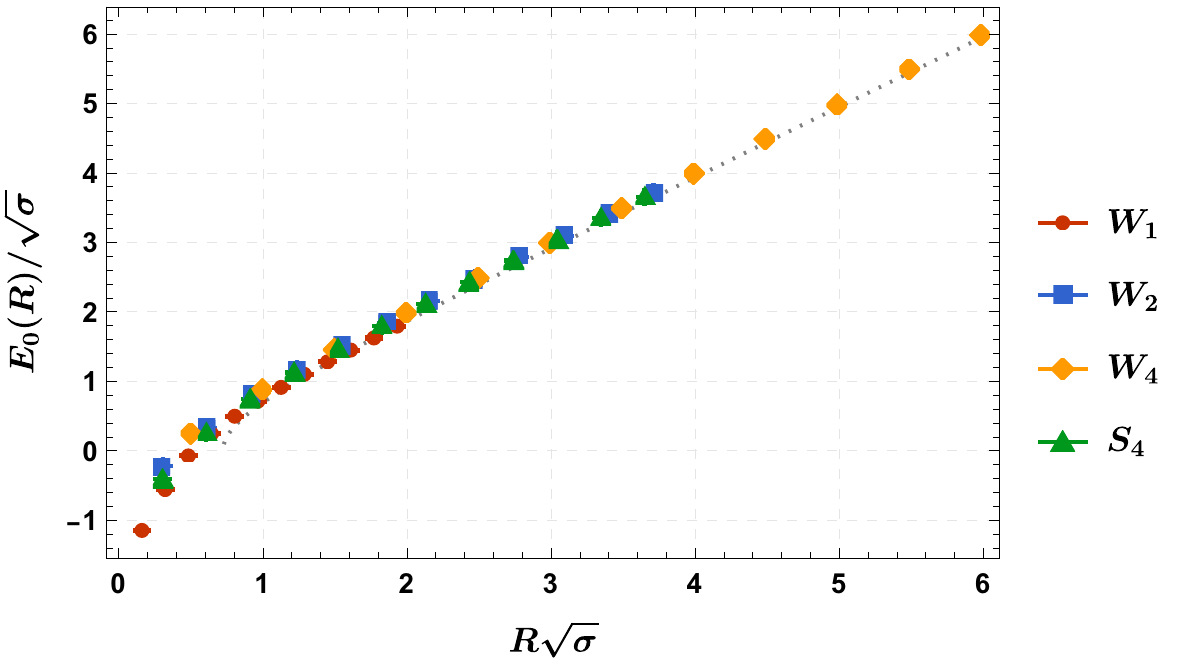}
\includegraphics[width=\columnwidth]{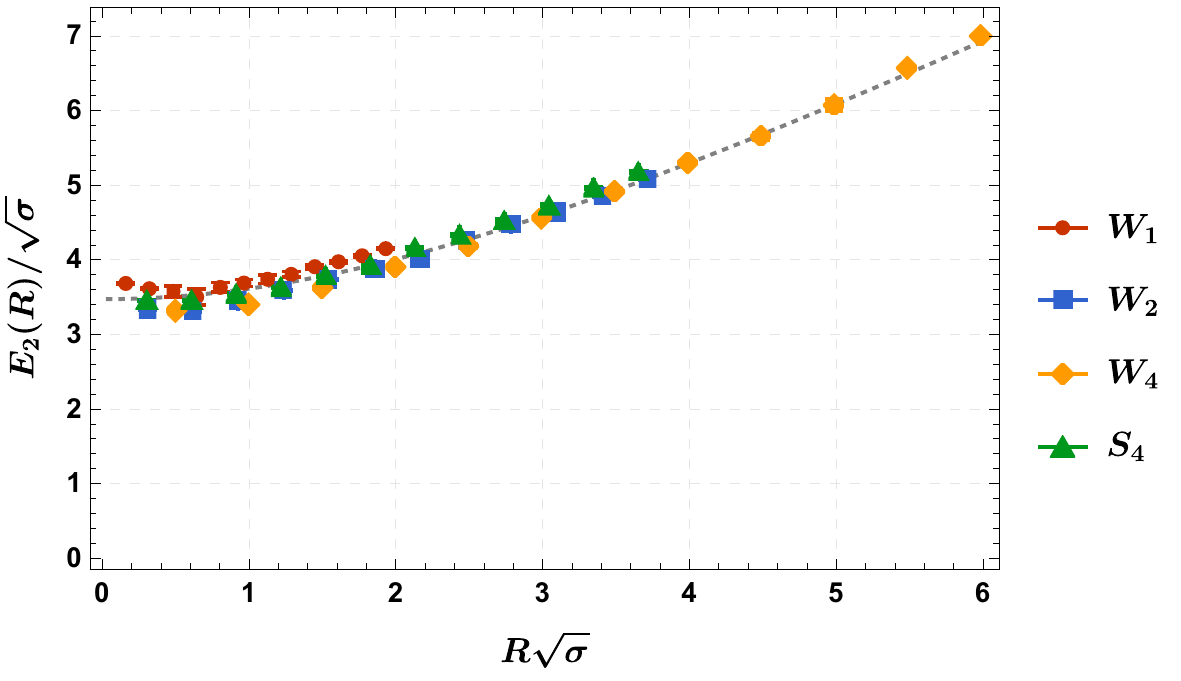}
\includegraphics[width=\columnwidth]{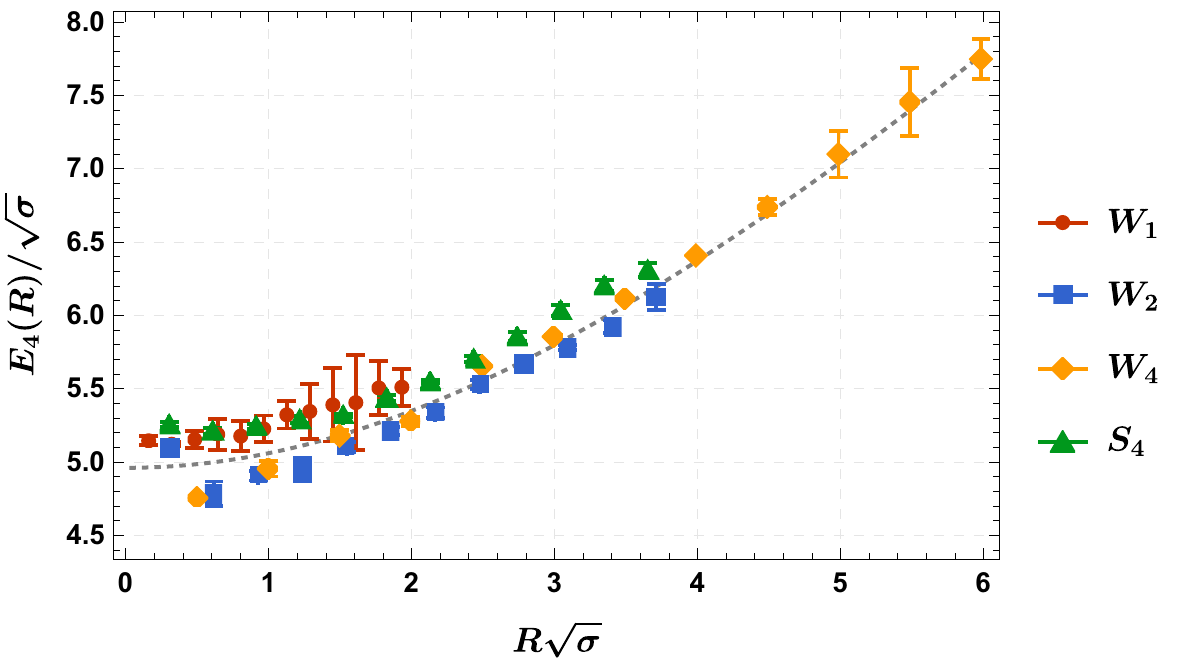}
\includegraphics[width=\columnwidth]{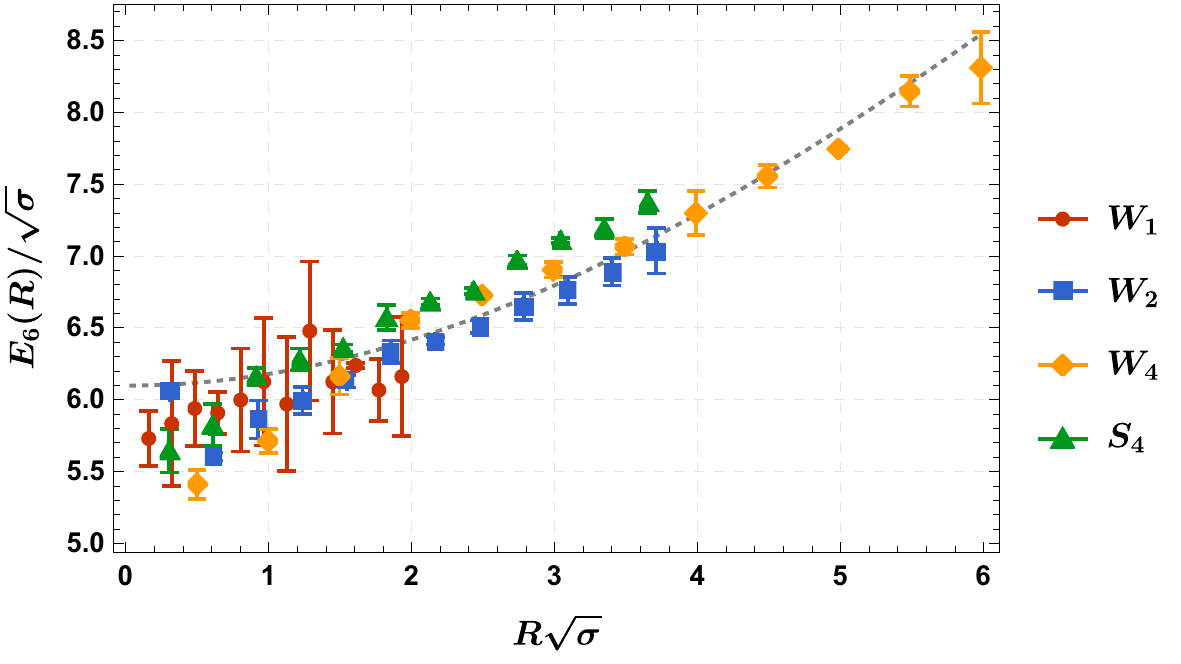}
\includegraphics[width=\columnwidth]{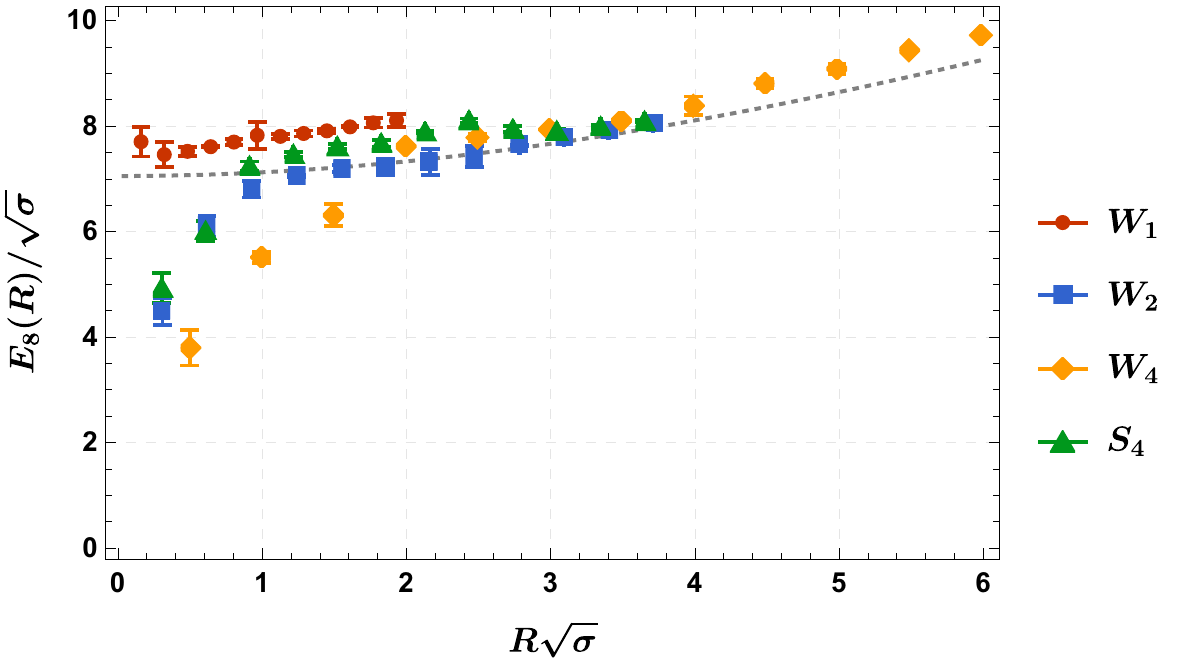}
\includegraphics[width=\columnwidth]{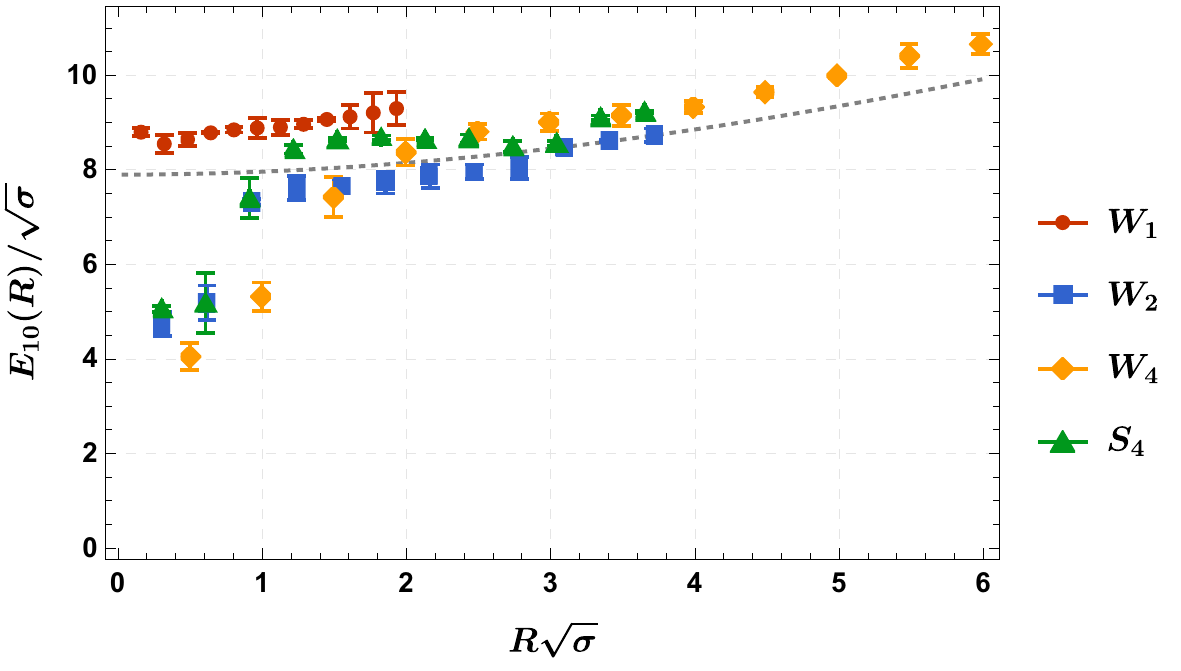}
\par\end{centering}
\caption{Comparing the results of ensembles $W_1$, $W_2$, $W_4$ and $S_4$ for each one of the energy levels. From left to right and top to bottom we show the groundstate potential and the first five excited states. To guide the eye, the dashed lines correspond to the corresponding level of the Nambu-Goto spectrum.}
\label{fig:pot_eachlevel}
\end{figure*}

We also compare the results of all actions in Fig. \ref{fig:pot_eachlevel}, where we study one potential level in each panel. The excited levels tend to be higher than the Nambu-Goto model. 

To conclude on the different spectra, we are able to get several levels, with on-axis operators, with smearing, and with anisotropic lattices. But we should discard the first four smaller distances for the anisotropic action because the levels above $N=2$ are distorted and tend to get degenerate.

%SSSSSSSSSSSSSSSSSSSSSSSSSSSSSSSSSSSSSSSSSSSSSSSSSSSSSSSSSSSSSSSSSSSSSSSS
%SSSSSSSSSSSSSSSSSSSSSSSSSSSSSSSSSSSSSSSSSSSSSSSSSSSSSSSSSSSSSSSSSSSSSSSS
%SSSSSSSSSSSSSSSSSSSSSSSSSSSSSSSSSSSSSSSSSSSSSSSSSSSSSSSSSSSSSSSSSSSSSSSS
\section{Analysis of our results \label{sec:analysis}}

The simplest description of the QCD excited flux tubes - with charges in the triplet representation of SU(3) - is given by the bosonic string
model, based on the Nambu-Goto action
\cite{Nambu:1978bd,Goto:1971ce},
\be
S=-\sigma\int d^{2}\Sigma \ ,
\ee
which spectrum for an open string with ends fixed with Dirichlet boundary conditions is the Arvis potential  \cite{Arvis:1983fp},
\be
V_{n}(R)=\sigma\sqrt{R^{2}+\frac{2\pi}{\sigma}\left(n-\frac{D-2}{24}\right)}\ .
\label{Arvis}
\ee
In Eq. (\ref{Arvis}), $D$ is the dimension of the space time $D=4$, and $n= 2 \, n_r+l$ is the principal quantum number. In our case of $\Sigma_g^+$, $l=0$ and the only quantum number we have is $n=2 \, n_r$ where $n_r$ is the order of the radial excitation. Because of this simple analytical form, we opt to fit the excited spectrum with the Arvis potential. Any deviation may indicate other phenomena, say a constituent gluon.

Note that, for the groundstate,  the Arvis potential is tachyonic at small distances since the argument of the square root is negative, moreover rotational invariance is only achieved for $D=26$.
Nevertheless  the first two terms in the $1/R$ expansion, including the Coulomb term,  are more general that the Arvis potential, since they fit the $D=3$ and $D=4$ lattice data quite well beyond the tachyonic distance. 
The Coulomb term is independent of the string tension $\sigma$ and for the physical
$D=3+1$ has the value $-\frac{\pi}{12} \frac{1}{R}$. This is the L\"uscher term  \cite{Luscher:1980iy}.
The energy spectrum of a static quark-antiquark and of its flux tube is certainly well defined (not tachionic) and this was the first evidence of flux tube vibrations found in lattice field theory.

In what concerns the groundstate, is well known, as detailed in Section \ref{sec:intro} that the Nambu-Goto fails, because we have no tachyon at short distances. However, the leading terms in a large distance expansion are accurate not only at long distances, but also at medium range short distances, where the tachyon is replaced by the L\"uscher \cite{Luscher:1980iy} coulombic potential,
\be
V_0(R)=\sigma R
- \frac{\pi}{12} \frac{1}{R}
+  O\left( {1 \over R^3} \right)
\ee
which confirms the factor $\left(n-\frac{D-2}{24}\right)$ in the Arvis potential. At very short distances the correct potential matches perturbative QCD \cite{Karbstein:2014bsa}. 

Nevertheless, for the excited states  the intrinsic width of the flux tube \cite{Cardoso:2013lla} should become negligible compared with the quantum vibrations of the string. The Nambu-Goto model should then be adequate to analyse the potentials we compute with lattice QCD. Notice the Arvis potential produces a tower of levels, 
\bea
%\sqrt{ 2\pi \sigma\left(n-\frac{D-2}{24}\right)} \left[1 +\frac{\sigma}{4 \pi \left(n- {D-2 \over 24} \right)} R^2 +  O\left( { R^4} \right) \right]
V_{n_r}(R) &=& \sigma \sqrt{ R^2 + {2\pi \over \sigma}
\left(2 \, n_r -\frac{1}{12}\right)}  
\non \\
&=& \sqrt{ 2\pi \sigma\left(2 \, n_r -\frac{1}{12}\right)}  +  O\left( { R^2} \right),
\label{NGex}
\eea
as we observe in our lattice QCD data.

\begin{figure}[t!]
\begin{centering}
\includegraphics[width=0.7\columnwidth]{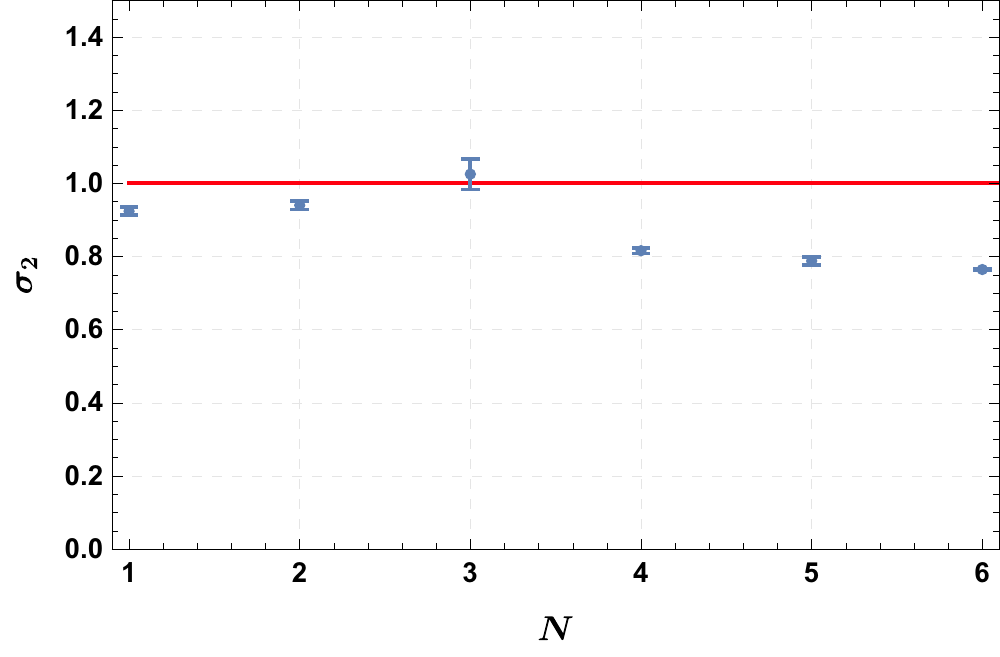}
\includegraphics[width=0.7\columnwidth]{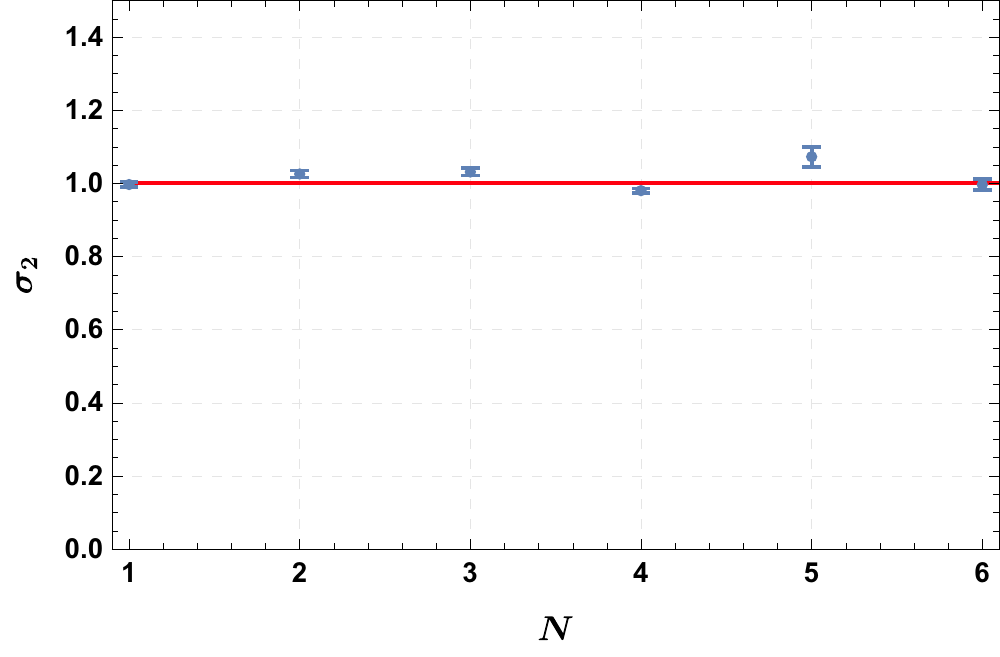}
\includegraphics[width=0.7\columnwidth]{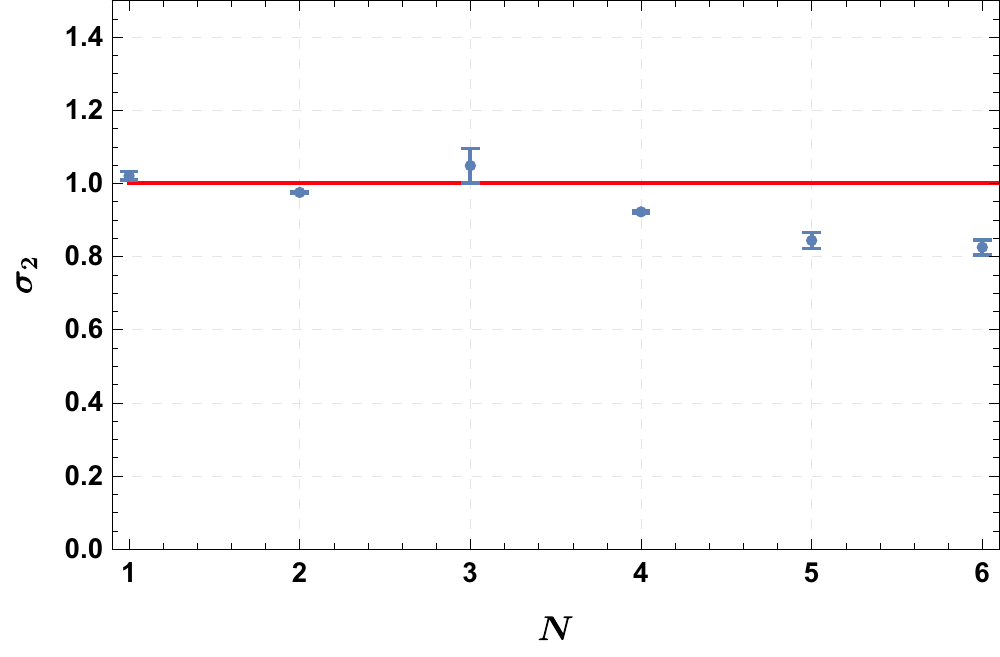}
\includegraphics[width=0.7\columnwidth]{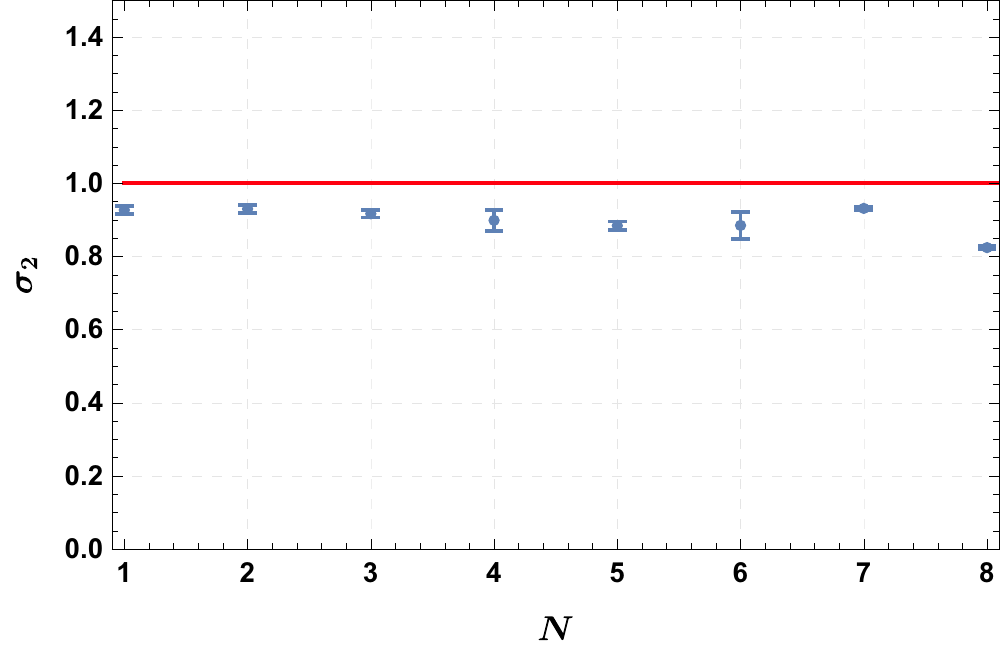}
\par\end{centering}
\caption{Results of the fits, for the second $\sigma_2$ parameter of the modified Nambu-Goto expression,   from top to bottom respectively with ensembles $W_1$, $W_2$, $W_4$ and $S_4$.}
\label{fig:pot_fits0}
\end{figure}

%--------------
\begin{figure}[t!]
\begin{centering}
\includegraphics[width=\columnwidth]{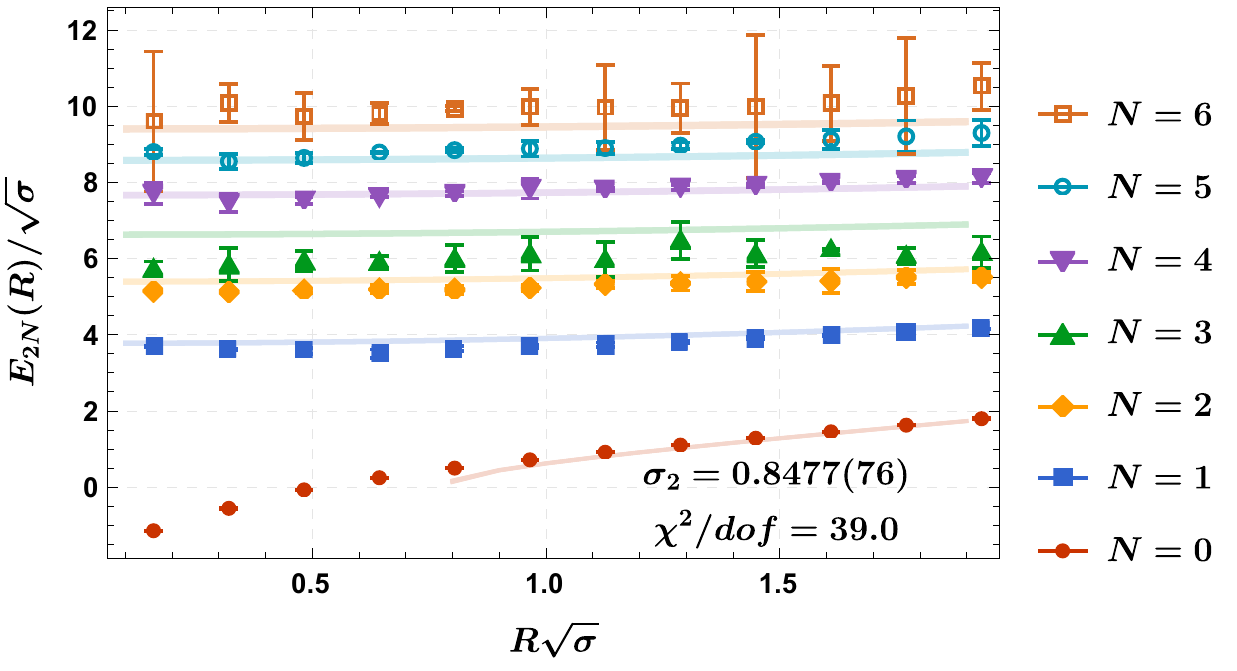}
\includegraphics[width=\columnwidth]{{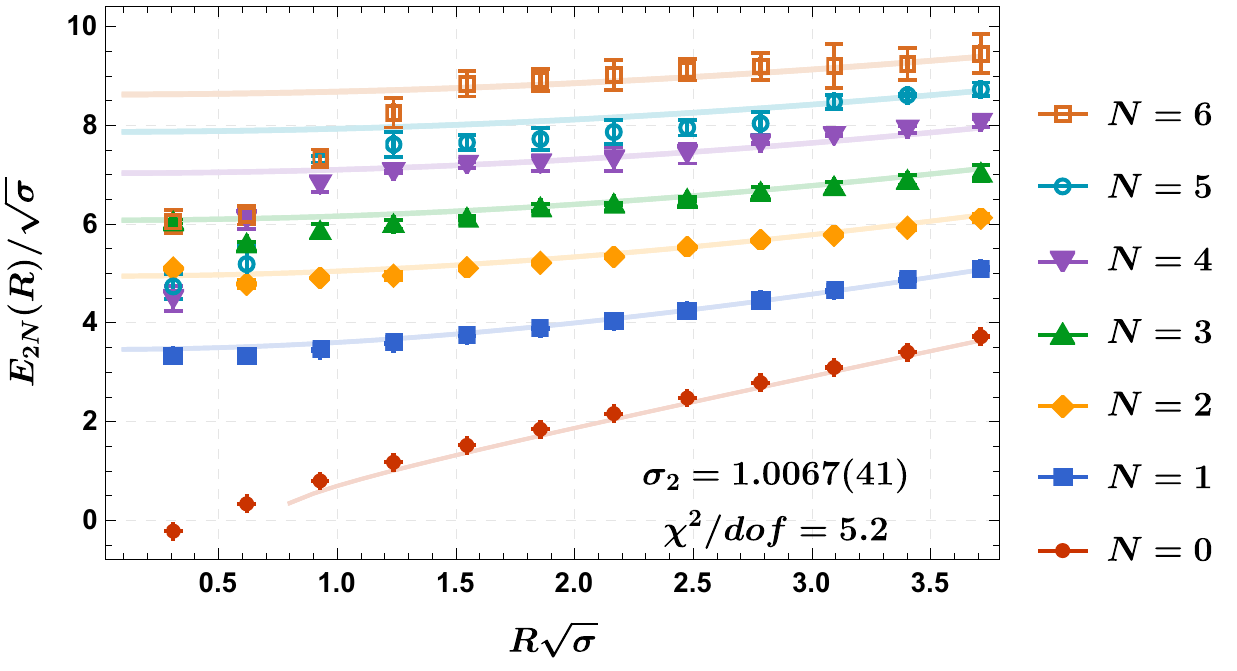}}
\includegraphics[width=\columnwidth]{{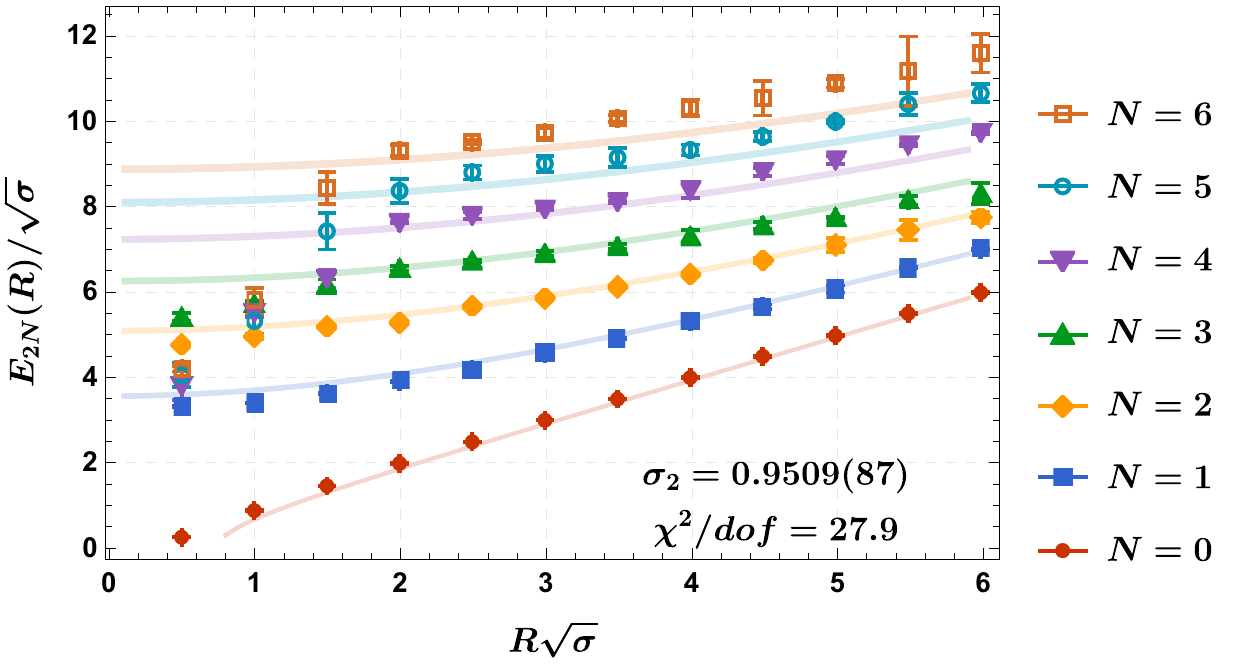}}
\includegraphics[width=\columnwidth]{{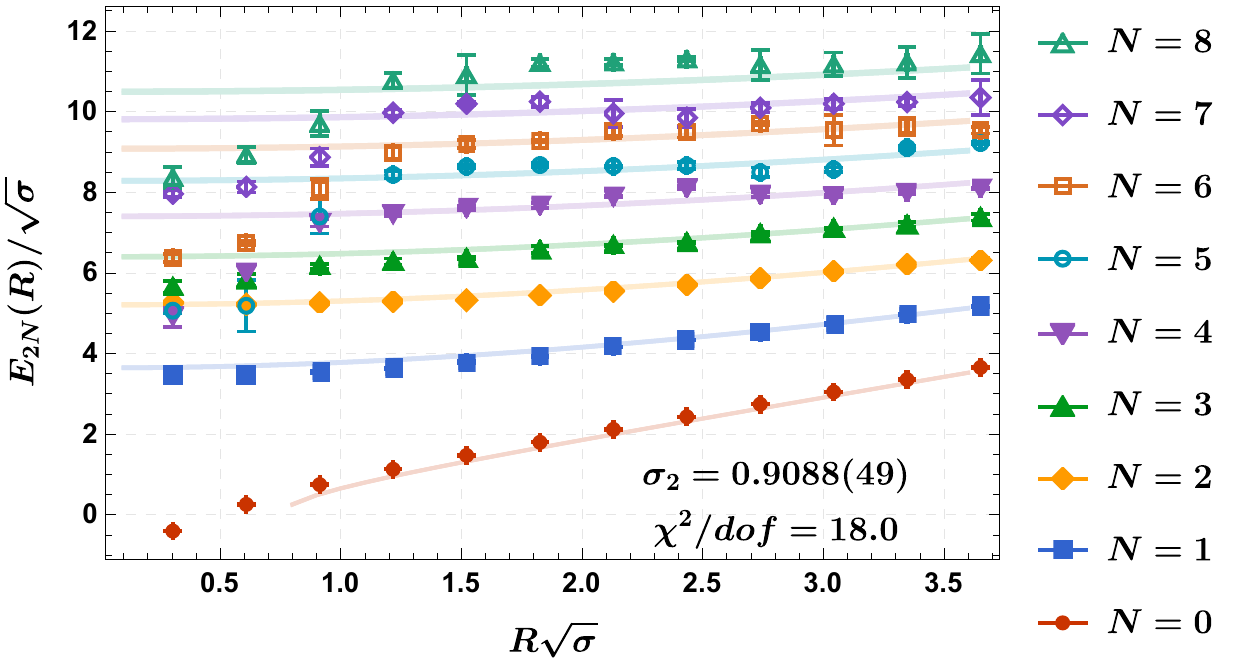}}
\par\end{centering}
\caption{Comparing our results to a fit with the modified Nambu-Gotto action, a unique $\sigma_2$ is extracted from a nonlinear multifit of all the energy levels, in the interval $[5a_s-12a_s]$,
from top to bottom respectively with ensembles $W_1$, $W_2$, $W_4$ and $S_4$. The width of the solid lines are equal to the error bar of our fit.
\label{fig:pot_compara}}
\end{figure}

%--------------
\begin{figure}[t!]
\begin{centering}
\includegraphics[width=\columnwidth]{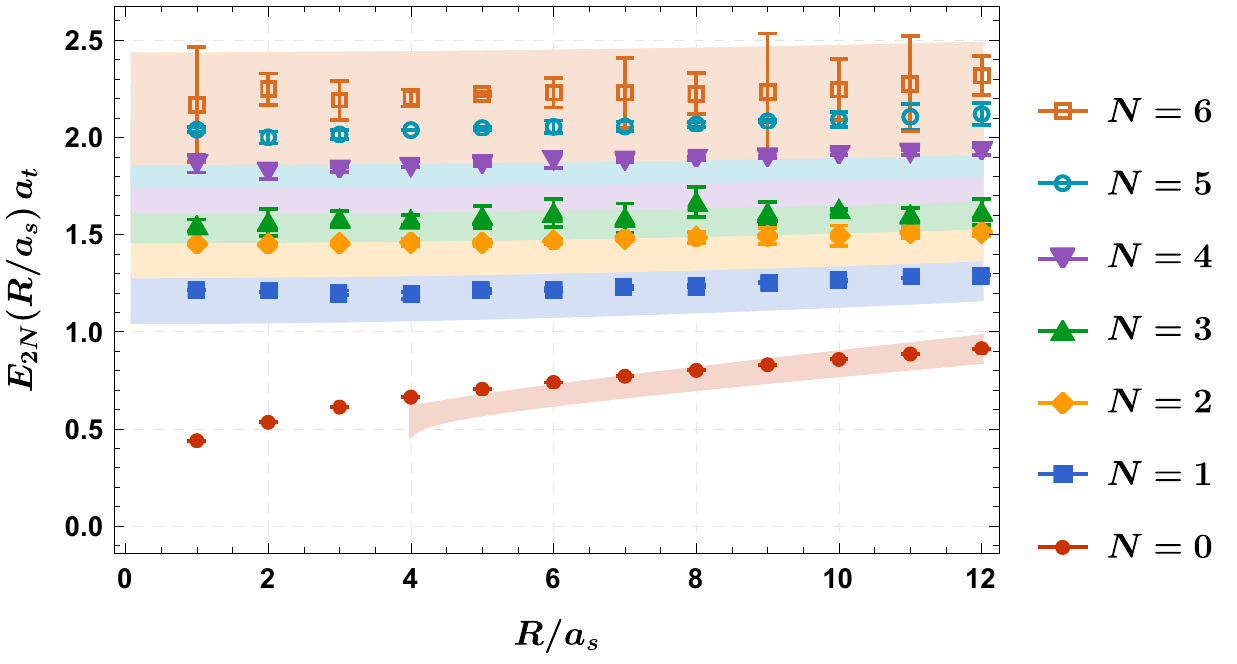}
\includegraphics[width=\columnwidth]{{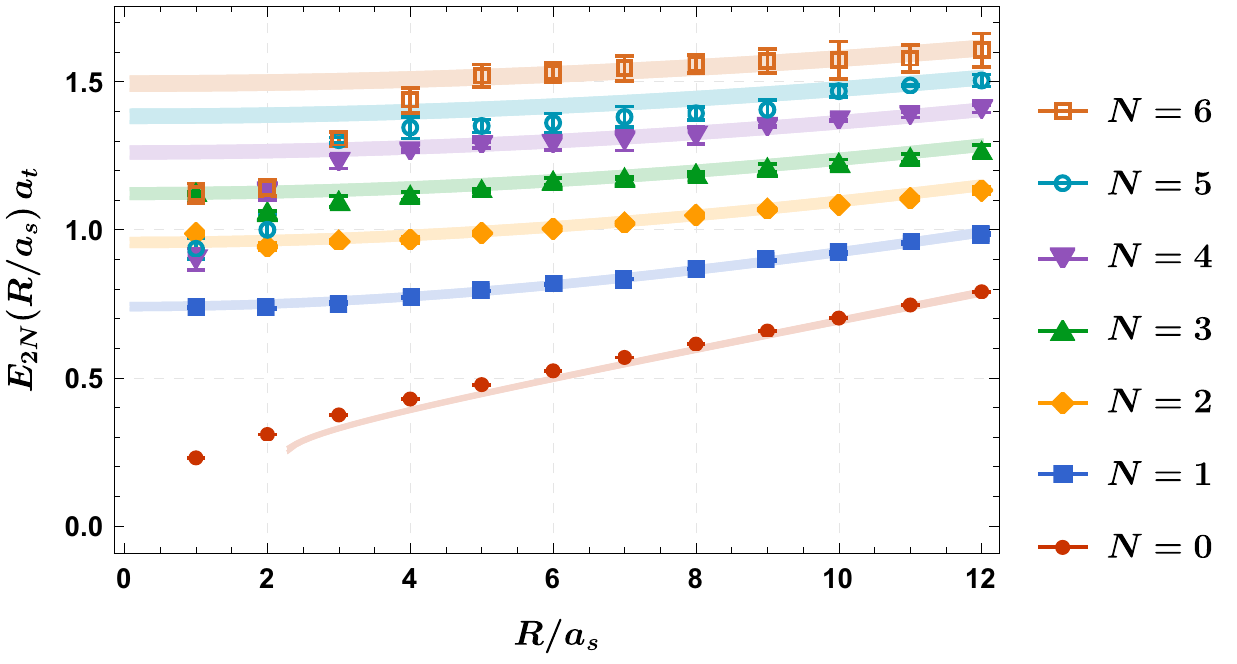}}
\includegraphics[width=\columnwidth]{{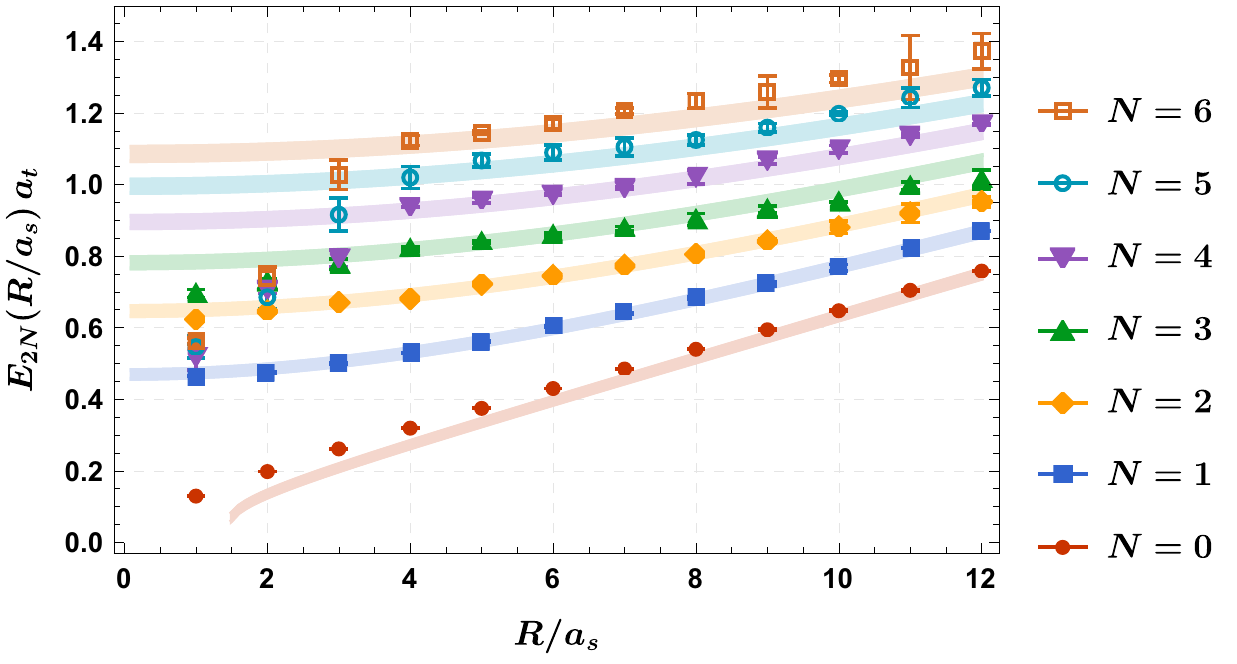}}
\includegraphics[width=\columnwidth]{{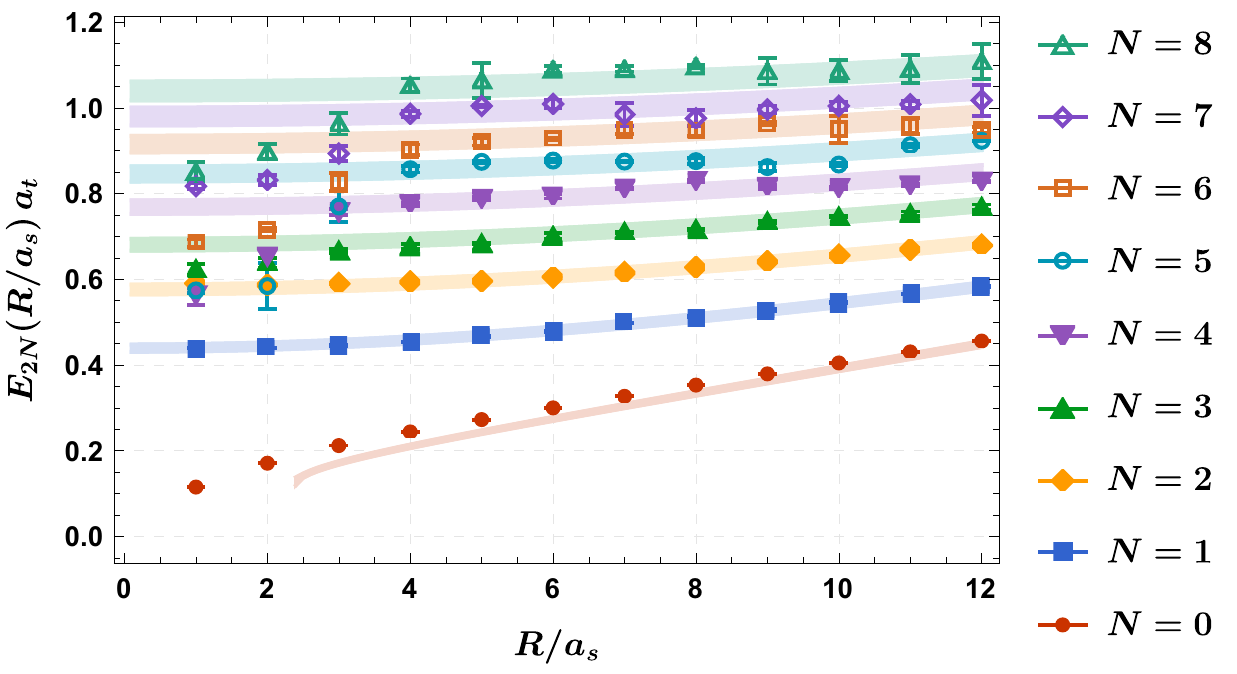}}
\par\end{centering}
\caption{Comparing our results in lattice spacing units to a fit with the modified Nambu-Gotto action the values are extracted from a nonlinear multifit of all the energy levels, in the interval $[5a_s-12a_s]$,
from top to bottom respectively with ensembles $W_1$, $W_2$, $W_4$ and $S_4$. The width of the solid lines are equal to the error bar of our fit. The groundstate was excluded from the multifit. 
$\sigma$ is the linear term of the groundstate fitted alone.
\label{fig:pot_compara1}}
\end{figure}

We now start analysing our data, shown in Figs. \ref{fig:pot_opera_1244} and \ref{fig:pot_eachlevel}, where they are compared explicitly with the Nambu-Goto model. 

\textcolor{black}{
If we exclude the smaller four distances showing some 
energy degeneracy, evident in Fig.  \ref{fig:pot_opera_1244} in the compression of the higher levels with $N > 2$, 
we see no evidence for the $\Gamma_g$ states, discussed in Subsection \ref{sec:operator}, degenerate with our $\Sigma_g^+$ states. 
Notice the degeneracy in these smaller four distances is possibly due to higher harmonics of the string vibrations. The harmonics with an even number of nodes have the same quantum numbers  $\Sigma_g^+$ as the fundamental harmonic. Since our operators only have two nodes, at the open ends of the string, they should not have a large overlap with the higher harmonics, except at shorter distances where the square shape of our operators may be too crude to only overlap with the fundamental harmonic. Since we are only interested in studying the radial excitations, From now on, we will exclude the four shortest distances from the fits of our data.
}

Moreover, we see no evidence for string breaking \cite{Bali:2005fu,Bulava:2019iut,Bicudo:2019ymo}, noticing with the anisotropic actions we are able to go to distances much larger than the string breaking distance. Thus we are confident to be analysing the pure hybrid  $\Sigma_g^+$ states. 

The first two excited states seem to be in general compatible with the Nambu-Goto model within error bars. However, the next states depart from it, apparently the energy levels are in general higher than in Nambu-Goto.

Thus, we analyse quantitatively the difference to the Nambu-Goto potential, fitting the data with an extra parameter. We multiply the constant term in the square root by an independent prefactor. The simplest way we find to do this is to change $\sigma$ in to $\sigma_2$ in the denominator or the constant term depending on $n_r$. The modified Nambu-Goto model is,
\be
V_{n}(R) = \sigma\,R\sqrt{1+\frac{2\pi}{\sigma_2 R^{2}}\left(n-\frac{D-2}{24}\right)}\ .
\label{modified}
\ee
In particular we fit the results of Fig. \ref{fig:pot_opera_1244} for each excited state separately with $E_{2N}(R)/(R\sigma)$ to $\sqrt{1 + \frac{2 \pi}{\sigma_2 R^2}\left( 2N - \frac{1}{12}\right)}$ and extract $\sigma_2$.
%$E^2_{2N}(R)/(R^2\sigma^2)$ to a constant $1 + \frac{1}{\sigma_2 R^2}\left(2 \pi (2N - \frac{1}{12}\right)$ and extract $\sigma_2$, propagating as well the error bars.
The resulting fits of all the energy levels for the different ensembles are shown in Fig. \ref{fig:pot_fits0}, and detailed in Table \ref{tab:fitsigma2}.

%--------------
\begin{table}[t!]
\begin{ruledtabular}
\begin{tabular}{c|cccc}
energy level  &  & \multicolumn{2}{c}{$\sigma_2/\sigma$} &   \\
  & $W_1$ & $W_2$ & $W_4$ & $S_4$  \\
\hline
$N_1$  &  0.9241(105)  &  0.9970(76)  &  1.0206(117)  &  0.9273(101)  \\
$N_2$  &  0.9403(114)  &  1.0258(93)  &  0.9753(20)  &  0.9302(112)  \\
$N_3$  &  1.0255(422)  &  1.0313(99)  &  1.0486(468)  &  0.9171(103)  \\
$N_4$  &  0.8160(79)  &  0.9803(58)  &  0.9223(37)  &  0.8990(285)  \\
$N_5$  &  0.7880(115)  &  1.0731(275)  &  0.8440(219)  &  0.8845(117)  \\
$N_6$  &  0.7646(14)  &  0.9967(147)  &  0.8254(203)  &  0.8852(363)  \\
$N_7$  &  -  &  -  &  -  &  0.9317(38)  \\
$N_8$  &  -  &  -  &  -  &  0.8245(46)  \\
\end{tabular}
\end{ruledtabular}
\caption{The different parameters $\sigma_2/\sigma$ in the modified Nambu-Goto model fitted for all the energy levels in the different ensembles.
}
\label{tab:fitsigma2}
\end{table}

Besides, we also use another approach with a gobal fit of all levels. Instead of fitting each excited state separately, we perform a non-linear multifit for all the excited states in each ensemble in Fig. \ref{fig:pot_compara}. For each emsemble, we display the resulted $\sigma_2$ and $\chi^2$/dof.
%Moreover, to visualize the quality of the fits, and display the $\chi^2$/dof test, we show the points and the adjusted curves in Fig. \ref{fig:pot_compara}. \textcolor{black}{XXX  non-linear multifit needs to be explained  XXX}
We find that for ensembles $W_1$, $W_4$ and $S_4$, $\sigma_2$ is up to 10 \% smaller than $\sigma$ (meaning that the energy levels are higher than in the Nambu-Goto model), but for $W_2$ and $\sigma_2$ are similar.

%--------------
\begin{table*}[t!]
\begin{ruledtabular}
\begin{tabular}{cccccccc}
ensemble  & $a_0$ & $\sigma$  & $\sigma_1$ & $\sigma_2$ & $\sigma_2/\xi_R$ & $\sigma_2/\xi_R/\sigma_1$ & $\chi^2/dof$ \\
\hline
$W_1$	& 0.521(408) & - & 0.0351(46) & 0.0344(74) & 0.0344(74) & 0.9814(2470) & 35.1 \\
$W_2$	& 0.2363(33) & 0.0440(1) & 0.0468(6) & 0.1039(25) & 0.0478(11) & 1.0205(270) & 2.9 \\
$W_4$	& 0.0564(99) & 0.0547(1) & 0.0583(8) & 0.2400(47) & 0.0528(10) & 0.9062(216) & 25.8 \\
$S_4$	& 0.1203(39) & 0.0255(1) & 0.0279(5) & 0.0923(29) & 0.0255(8) & 0.9116(329) & 13.7 \\
\end{tabular}
\end{ruledtabular}
\caption{Fit results of the Fig. \ref{fig:pot_compara1} extracted from a nonlinear multifit of all the energy levels, in the interval $[5a_s-12a_s]$. The groundstate was excluded from the multifit. $\sigma$ is the linear term of the groundstate fitted alone.
%Our ensembles, for the isotropic Wilson action and the improved anisotropic $S_{II}$ action. $\xi$ is the bare anisotropy in the Lagrangian and $\xi_R$ is the renormalized anisotropy. The renormalized anisotropy and the lattice spacings are computed with the prescription of Section \ref{sec:renormalized}.
}
\label{tab:fig11data}
\end{table*}

Furthermore, since the results in Fig. \ref{fig:pot_opera_1244} are in units of the string tension obtained from the groundstate fit, we also perform an even more global fit, fitting all parameters to the excited spectrum (and not fitting the groundstate). In Fig. \ref{fig:pot_compara1} and in Table \ref{tab:fig11data}  we depict the non-linear multifit for all the excited states in lattice spacing units to the equation: 
\be
V_{n}(R) = a_0 + \sigma_1\,R\sqrt{1+\frac{2\pi}{\sigma_2 R^{2}}\left(n-\frac{D-2}{24}\right)}.
\label{modified1}
\ee
In this case what we have to compare with the previous $\sigma_2 / \sigma$ is now $\sigma_2 / \xi_R /  \sigma_1$. Again this goes down, up to 10\% in ensembles $W_4$ and $S_4$, while in ensembles $W_1$ and $W_2$ this is of order of 1.

This departure of up to 10 \% from the Nambu-Goto model, may possibly be interpreted as the string being non-homogeneous, or be interpreted with the existence of a constituent gluon in the more excited states.

%SSSSSSSSSSSSSSSSSSSSSSSSSSSSSSSSSSSSSSSSSSSSSSSSSSSSSSSSSSSSSSSSSSSSSSSS
%SSSSSSSSSSSSSSSSSSSSSSSSSSSSSSSSSSSSSSSSSSSSSSSSSSSSSSSSSSSSSSSSSSSSSSSS
%SSSSSSSSSSSSSSSSSSSSSSSSSSSSSSSSSSSSSSSSSSSSSSSSSSSSSSSSSSSSSSSSSSSSSSSS
\section{Conclusions and outlook \label{sec:conclu}}

We computed the potentials for several new excitations of the pure $SU(3)$ flux tubes produced by two static $3$ and $\bar 3$ sources, specializing in the radial excitations of the representation $\Sigma^+_g$. 

We succeeded in obtaining the spectrum of several new excited flux tubes, using a large basis of operators , employing the computational techniques with GPUs of Ref. \cite{Bicudo:2018jbb}  (improved to be able to study field densities), and utilizing different actions with smearing and anisotropy. Previously in the literature and websites, states up to $N=2$ have been shown, and we go up to $N=8$.

In general the excited states of the the $\Sigma^+_g$ flux tubes are comparable to the Nambu-Goto string model with transverse modes, only depending on the string tension $\sigma$ and the radial quantum number $n_r$.

Nevertheless we find evidence that the $\Sigma^+_g$ flux tubes cannot be exactly modelled by the Nambu-Goto string model. A second parameter, we parametrize as a second $\sigma_2$ up to 10 \% smaller than $\sigma$, essentially corresponding to a larger energy splitting between levels than in the Nambu-Goto model leads to better fits. Interesting for the theoretical studies of the QCD flux tubes, this may indicate the formation of a non-homogenous string with a lump or a constituent gluon, or of an extra symmetry higher in the spectrum as referred in Section \ref{sec:intro}. 

To clarify in more detail this small  tension with the Nambu-Goto model, we would need to have more computational power, to be able to use larger lattices and more operators. Another direction of research is on investigating higher angular excitations of the flux tube (here we investigated radial excitations), but a new approach would be necessary to overcome the limitations of a cubic lattice. 
\textcolor{black}{
Searching in different quantum numbers for explicit evidences of a constituent gluon \cite{Mueller:2019mkh} or of an axion \cite{Dubovsky:2013gi} is also motivating.
}
We leave this for future works.

\vspace{0.25cm}
\textbf{Acknowledgments}
\vspace{0.1cm}

We acknowledge discussions on flux tubes and constituent gluons with our colleagues Sergey Afonin, Gunnar Bali, Daniele Binosi,  Bastian Brandt, Richard Brower, Fabien Buisseret, Leonid Glozman, Alexei Kaidalov, Felipe Llanes-Estrada, Vincent Mathieu, Colin Morningstar, Lasse Müller, Fumiko Okiharu, Orlando Oliveira, Emilio Ribeiro and Marc Wagner.
We thank the support of
CeFEMA under the contract for R\&D Units, strategic project No. UID/CTM/04540/2019, and the FCT project Grant No. CERN/FIS-COM/0029/2017.
NC is supported by 
FCT under the contract No. SFRH/BPD/109443/2015 and 
AS is supported by the contract No. SFRH/PD/BD/135189/2017.

\bibliographystyle{elsarticle-num}
\bibliography{excited}{}

\begin{thebibliography}{10}
\expandafter\ifx\csname url\endcsname\relax
  \def\url#1{\texttt{#1}}\fi
\expandafter\ifx\csname urlprefix\endcsname\relax\def\urlprefix{URL }\fi
\expandafter\ifx\csname href\endcsname\relax
  \def\href#1#2{#2} \def\path#1{#1}\fi

\bibitem{Wilson:1974sk}
K.~G. Wilson, {Confinement of Quarks}, Phys. Rev. D10 (1974) 2445--2459,
  [,45(1974)].
\newblock \href {https://doi.org/10.1103/PhysRevD.10.2445}
  {\path{doi:10.1103/PhysRevD.10.2445}}.

\bibitem{Bicudo:2007wt}
P.~Bicudo, {The large degeneracy of excited hadrons and quark models}, Phys.
  Rev. D76 (2007) 094005.
\newblock \href {http://arxiv.org/abs/hep-ph/0703114}
  {\path{arXiv:hep-ph/0703114}}, \href
  {https://doi.org/10.1103/PhysRevD.76.094005}
  {\path{doi:10.1103/PhysRevD.76.094005}}.

\bibitem{Bicudo:2009hm}
P.~Bicudo, {Gluon Excitations and Quark Chiral Symmetry in the Meson Spectrum:
  An Einbein Solution to the Large Degeneracy Problem of Light Mesons}, Phys.
  Rev. D 81 (2010) 014011.
\newblock \href {http://arxiv.org/abs/0904.0030} {\path{arXiv:0904.0030}},
  \href {https://doi.org/10.1103/PhysRevD.81.014011}
  {\path{doi:10.1103/PhysRevD.81.014011}}.

\bibitem{Godfrey:1985xj}
S.~Godfrey, N.~Isgur, {Mesons in a Relativized Quark Model with
  Chromodynamics}, Phys. Rev. D32 (1985) 189--231.
\newblock \href {https://doi.org/10.1103/PhysRevD.32.189}
  {\path{doi:10.1103/PhysRevD.32.189}}.

\bibitem{Isgur:1978xj}
N.~Isgur, G.~Karl, {P Wave Baryons in the Quark Model}, Phys. Rev. D18 (1978)
  4187.
\newblock \href {https://doi.org/10.1103/PhysRevD.18.4187}
  {\path{doi:10.1103/PhysRevD.18.4187}}.

\bibitem{Nambu:1978bd}
Y.~Nambu, {QCD and the String Model}, Phys. Lett. B 80 (1979) 372--376.
\newblock \href {https://doi.org/10.1016/0370-2693(79)91193-6}
  {\path{doi:10.1016/0370-2693(79)91193-6}}.

\bibitem{Goto:1971ce}
T.~Goto, {Relativistic quantum mechanics of one-dimensional mechanical
  continuum and subsidiary condition of dual resonance model}, Prog. Theor.
  Phys. 46 (1971) 1560--1569.
\newblock \href {https://doi.org/10.1143/PTP.46.1560}
  {\path{doi:10.1143/PTP.46.1560}}.

\bibitem{Aharony:2009gg}
O.~Aharony, E.~Karzbrun, {On the effective action of confining strings}, JHEP
  06 (2009) 012.
\newblock \href {http://arxiv.org/abs/0903.1927} {\path{arXiv:0903.1927}},
  \href {https://doi.org/10.1088/1126-6708/2009/06/012}
  {\path{doi:10.1088/1126-6708/2009/06/012}}.

\bibitem{Gliozzi:2010zv}
F.~Gliozzi, M.~Pepe, U.~J. Wiese, {The Width of the Confining String in
  Yang-Mills Theory}, Phys. Rev. Lett. 104 (2010) 232001.
\newblock \href {http://arxiv.org/abs/1002.4888} {\path{arXiv:1002.4888}},
  \href {https://doi.org/10.1103/PhysRevLett.104.232001}
  {\path{doi:10.1103/PhysRevLett.104.232001}}.

\bibitem{Alvarez:1981kc}
O.~Alvarez, {The Static Potential in String Models}, Phys. Rev. D24 (1981) 440.
\newblock \href {https://doi.org/10.1103/PhysRevD.24.440}
  {\path{doi:10.1103/PhysRevD.24.440}}.

\bibitem{Arvis:1983fp}
J.~F. Arvis, {The Exact $q \bar{q}$ Potential in Nambu String Theory}, Phys.
  Lett. 127B (1983) 106--108.
\newblock \href {https://doi.org/10.1016/0370-2693(83)91640-4}
  {\path{doi:10.1016/0370-2693(83)91640-4}}.

\bibitem{Karbstein:2014bsa}
F.~Karbstein, A.~Peters, M.~Wagner,
  {${\Lambda}_{\overline{\mathrm{MS}}}^{({n}_f=2)}$ from a momentum space
  analysis of the quark-antiquark static potential}, JHEP 09 (2014) 114.
\newblock \href {http://arxiv.org/abs/1407.7503} {\path{arXiv:1407.7503}},
  \href {https://doi.org/10.1007/JHEP09(2014)114}
  {\path{doi:10.1007/JHEP09(2014)114}}.

\bibitem{PtQCD}
\url{http://saturno.ist.utl.pt/~ptqcd} (2007).

\bibitem{Cardoso:2013lla}
N.~Cardoso, M.~Cardoso, P.~Bicudo, {Inside the SU(3) quark-antiquark QCD flux
  tube: screening versus quantum widening}, Phys. Rev. D88 (2013) 054504.
\newblock \href {http://arxiv.org/abs/1302.3633} {\path{arXiv:1302.3633}},
  \href {https://doi.org/10.1103/PhysRevD.88.054504}
  {\path{doi:10.1103/PhysRevD.88.054504}}.

\bibitem{Bugg:2004xu}
D.~V. Bugg, {Four sorts of meson}, Phys. Rept. 397 (2004) 257--358.
\newblock \href {http://arxiv.org/abs/hep-ex/0412045}
  {\path{arXiv:hep-ex/0412045}}, \href
  {https://doi.org/10.1016/j.physrep.2004.03.008}
  {\path{doi:10.1016/j.physrep.2004.03.008}}.

\bibitem{Aker:1992ny}
E.~Aker, et~al., {The Crystal Barrel spectrometer at LEAR}, Nucl. Instrum.
  Meth. A 321 (1992) 69--108.
\newblock \href {https://doi.org/10.1016/0168-9002(92)90379-I}
  {\path{doi:10.1016/0168-9002(92)90379-I}}.

\bibitem{Glozman:2007ek}
L.~Y. Glozman, {Restoration of chiral and U(1)A symmetries in excited hadrons},
  Phys. Rept. 444 (2007) 1--49.
\newblock \href {http://arxiv.org/abs/hep-ph/0701081}
  {\path{arXiv:hep-ph/0701081}}, \href
  {https://doi.org/10.1016/j.physrep.2007.04.001}
  {\path{doi:10.1016/j.physrep.2007.04.001}}.

\bibitem{Afonin:2006wt}
S.~S. Afonin, {Light meson spectrum and classical symmetries of QCD}, Eur.
  Phys. J. A 29 (2006) 327--335.
\newblock \href {http://arxiv.org/abs/hep-ph/0606310}
  {\path{arXiv:hep-ph/0606310}}, \href
  {https://doi.org/10.1140/epja/i2006-10080-2}
  {\path{doi:10.1140/epja/i2006-10080-2}}.

\bibitem{Afonin:2007jd}
S.~S. Afonin, {Towards understanding spectral degeneracies in nonstrange
  hadrons. Part I. Mesons as hadron strings versus phenomenology}, Mod. Phys.
  Lett. A 22 (2007) 1359--1372.
\newblock \href {http://arxiv.org/abs/hep-ph/0701089}
  {\path{arXiv:hep-ph/0701089}}, \href
  {https://doi.org/10.1142/S0217732307024024}
  {\path{doi:10.1142/S0217732307024024}}.

\bibitem{Glozman:2012fj}
L.~Y. Glozman, C.~B. Lang, M.~Schrock, {Symmetries of hadrons after unbreaking
  the chiral symmetry}, Phys. Rev. D 86 (2012) 014507.
\newblock \href {http://arxiv.org/abs/1205.4887} {\path{arXiv:1205.4887}},
  \href {https://doi.org/10.1103/PhysRevD.86.014507}
  {\path{doi:10.1103/PhysRevD.86.014507}}.

\bibitem{Catillo:2018cyv}
M.~Catillo, L.~Y. Glozman, {Baryon parity doublets and chiral spin symmetry},
  Phys. Rev. D 98~(1) (2018) 014030.
\newblock \href {http://arxiv.org/abs/1804.07171} {\path{arXiv:1804.07171}},
  \href {https://doi.org/10.1103/PhysRevD.98.014030}
  {\path{doi:10.1103/PhysRevD.98.014030}}.

\bibitem{Semay:2008nq}
C.~Semay, F.~Buisseret, B.~Silvestre-Brac, {Towers of hybrid mesons}, Phys.
  Rev. D 79 (2009) 094020.
\newblock \href {http://arxiv.org/abs/0812.3291} {\path{arXiv:0812.3291}},
  \href {https://doi.org/10.1103/PhysRevD.79.094020}
  {\path{doi:10.1103/PhysRevD.79.094020}}.

\bibitem{Abreu:2005uw}
E.~Abreu, P.~Bicudo, {Glueball and hybrid mass and decay with string tension
  below Casimir scaling}, J. Phys. G 34 (2007) 195207.
\newblock \href {http://arxiv.org/abs/hep-ph/0508281}
  {\path{arXiv:hep-ph/0508281}}, \href
  {https://doi.org/10.1088/0954-3899/34/2/003}
  {\path{doi:10.1088/0954-3899/34/2/003}}.

\bibitem{Buisseret:2006wc}
F.~Buisseret, C.~Semay, {On two- and three-body descriptions of hybrid mesons},
  Phys. Rev. D 74 (2006) 114018.
\newblock \href {http://arxiv.org/abs/hep-ph/0610132}
  {\path{arXiv:hep-ph/0610132}}, \href
  {https://doi.org/10.1103/PhysRevD.74.114018}
  {\path{doi:10.1103/PhysRevD.74.114018}}.

\bibitem{Cornwall:1981zr}
J.~M. Cornwall, {Dynamical Mass Generation in Continuum QCD}, Phys. Rev. D 26
  (1982) 1453.
\newblock \href {https://doi.org/10.1103/PhysRevD.26.1453}
  {\path{doi:10.1103/PhysRevD.26.1453}}.

\bibitem{Oliveira:2010xc}
O.~Oliveira, P.~Bicudo, {Running Gluon Mass from Landau Gauge Lattice QCD
  Propagator}, J. Phys. G 38 (2011) 045003.
\newblock \href {http://arxiv.org/abs/1002.4151} {\path{arXiv:1002.4151}},
  \href {https://doi.org/10.1088/0954-3899/38/4/045003}
  {\path{doi:10.1088/0954-3899/38/4/045003}}.

\bibitem{Bicudo:2007xp}
P.~Bicudo, M.~Cardoso, O.~Oliveira, {Study of the gluon-quark-antiquark static
  potential in SU(3) lattice QCD}, Phys. Rev. D 77 (2008) 091504.
\newblock \href {http://arxiv.org/abs/0704.2156} {\path{arXiv:0704.2156}},
  \href {https://doi.org/10.1103/PhysRevD.77.091504}
  {\path{doi:10.1103/PhysRevD.77.091504}}.

\bibitem{Cardoso:2009kz}
M.~Cardoso, N.~Cardoso, P.~Bicudo, {Lattice QCD computation of the colour
  fields for the static hybrid quark-gluon-antiquark system, and microscopic
  study of the Casimir scaling}, Phys. Rev. D 81 (2010) 034504.
\newblock \href {http://arxiv.org/abs/0912.3181} {\path{arXiv:0912.3181}},
  \href {https://doi.org/10.1103/PhysRevD.81.034504}
  {\path{doi:10.1103/PhysRevD.81.034504}}.

\bibitem{Dubovsky:2013gi}
S.~Dubovsky, R.~Flauger, V.~Gorbenko, {Evidence from Lattice Data for a New
  Particle on the Worldsheet of the QCD Flux Tube}, Phys. Rev. Lett. 111~(6)
  (2013) 062006.
\newblock \href {http://arxiv.org/abs/1301.2325} {\path{arXiv:1301.2325}},
  \href {https://doi.org/10.1103/PhysRevLett.111.062006}
  {\path{doi:10.1103/PhysRevLett.111.062006}}.

\bibitem{Athenodorou:2010cs}
A.~Athenodorou, B.~Bringoltz, M.~Teper, {Closed flux tubes and their string
  description in D=3+1 SU(N) gauge theories}, JHEP 02 (2011) 030.
\newblock \href {http://arxiv.org/abs/1007.4720} {\path{arXiv:1007.4720}},
  \href {https://doi.org/10.1007/JHEP02(2011)030}
  {\path{doi:10.1007/JHEP02(2011)030}}.

\bibitem{Lucini:2012gg}
B.~Lucini, M.~Panero, {SU(N) gauge theories at large N}, Phys. Rept. 526 (2013)
  93--163.
\newblock \href {http://arxiv.org/abs/1210.4997} {\path{arXiv:1210.4997}},
  \href {https://doi.org/10.1016/j.physrep.2013.01.001}
  {\path{doi:10.1016/j.physrep.2013.01.001}}.

\bibitem{Campbell:1987nv}
N.~A. Campbell, A.~Huntley, C.~Michael, {Heavy Quark Potentials and Hybrid
  Mesons From SU(3) Lattice Gauge Theory}, Nucl. Phys. B306 (1988) 51--62.
\newblock \href {https://doi.org/10.1016/0550-3213(88)90170-8}
  {\path{doi:10.1016/0550-3213(88)90170-8}}.

\bibitem{Perantonis:1990dy}
S.~Perantonis, C.~Michael, {Static potentials and hybrid mesons from pure SU(3)
  lattice gauge theory}, Nucl. Phys. B347 (1990) 854--868.
\newblock \href {https://doi.org/10.1016/0550-3213(90)90386-R}
  {\path{doi:10.1016/0550-3213(90)90386-R}}.

\bibitem{Lacock:1996ny}
P.~Lacock, C.~Michael, P.~Boyle, P.~Rowland, {Hybrid mesons from quenched QCD},
  Phys. Lett. B401 (1997) 308--312.
\newblock \href {http://arxiv.org/abs/hep-lat/9611011}
  {\path{arXiv:hep-lat/9611011}}, \href
  {https://doi.org/10.1016/S0370-2693(97)00384-5}
  {\path{doi:10.1016/S0370-2693(97)00384-5}}.

\bibitem{Lacock:1996vy}
P.~Lacock, C.~Michael, P.~Boyle, P.~Rowland, {Orbitally excited and hybrid
  mesons from the lattice}, Phys. Rev. D54 (1996) 6997--7009.
\newblock \href {http://arxiv.org/abs/hep-lat/9605025}
  {\path{arXiv:hep-lat/9605025}}, \href
  {https://doi.org/10.1103/PhysRevD.54.6997}
  {\path{doi:10.1103/PhysRevD.54.6997}}.

\bibitem{Juge:1999ie}
K.~J. Juge, J.~Kuti, C.~J. Morningstar, {Ab initio study of hybrid anti-b g b
  mesons}, Phys. Rev. Lett. 82 (1999) 4400--4403.
\newblock \href {http://arxiv.org/abs/hep-ph/9902336}
  {\path{arXiv:hep-ph/9902336}}, \href
  {https://doi.org/10.1103/PhysRevLett.82.4400}
  {\path{doi:10.1103/PhysRevLett.82.4400}}.

\bibitem{Juge:2002br}
K.~J. Juge, J.~Kuti, C.~Morningstar, {Fine structure of the QCD string
  spectrum}, Phys. Rev. Lett. 90 (2003) 161601.
\newblock \href {http://arxiv.org/abs/hep-lat/0207004}
  {\path{arXiv:hep-lat/0207004}}, \href
  {https://doi.org/10.1103/PhysRevLett.90.161601}
  {\path{doi:10.1103/PhysRevLett.90.161601}}.

\bibitem{Reisinger:2017btr}
C.~Reisinger, S.~Capitani, O.~Philipsen, M.~Wagner,
  \href{https://inspirehep.net/record/1616731/files/arXiv:1708.05562.pdf}{{Computation
  of hybrid static potentials in SU(3) lattice gauge theory}}, in: {35th
  International Symposium on Lattice Field Theory (Lattice 2017) Granada,
  Spain, June 18-24, 2017}, 2017.
\newblock \href {http://arxiv.org/abs/1708.05562} {\path{arXiv:1708.05562}}.
\newline\urlprefix\url{https://inspirehep.net/record/1616731/files/arXiv:1708.05562.pdf}

\bibitem{Mueller:2018fkg}
L.~Müller, M.~Wagner, {Structure of hybrid static potential flux tubes in
  SU(2) lattice Yang-Mills theory}, in: {10th International Winter Workshop
  "Excited QCD" 2018 Kopaonik, Serbia, March 11-15, 2018}, 2018.
\newblock \href {http://arxiv.org/abs/1803.11124} {\path{arXiv:1803.11124}}.

\bibitem{Bicudo:2018jbb}
P.~Bicudo, N.~Cardoso, M.~Cardoso, {Color field densities of the
  quark-antiquark excited flux tubes in SU(3) lattice QCD}, Phys. Rev. D
  98~(11) (2018) 114507.
\newblock \href {http://arxiv.org/abs/1808.08815} {\path{arXiv:1808.08815}},
  \href {https://doi.org/10.1103/PhysRevD.98.114507}
  {\path{doi:10.1103/PhysRevD.98.114507}}.

\bibitem{Mueller:2019mkh}
L.~M\"uller, O.~Philipsen, C.~Reisinger, M.~Wagner, {Hybrid static potential
  flux tubes from SU(2) and SU(3) lattice gauge theory}, Phys. Rev. D 100~(5)
  (2019) 054503.
\newblock \href {http://arxiv.org/abs/1907.01482} {\path{arXiv:1907.01482}},
  \href {https://doi.org/10.1103/PhysRevD.100.054503}
  {\path{doi:10.1103/PhysRevD.100.054503}}.

\bibitem{Capitani:2018rox}
S.~Capitani, O.~Philipsen, C.~Reisinger, C.~Riehl, M.~Wagner, {Precision
  computation of hybrid static potentials in SU(3) lattice gauge theory}, Phys.
  Rev. D 99~(3) (2019) 034502.
\newblock \href {http://arxiv.org/abs/1811.11046} {\path{arXiv:1811.11046}},
  \href {https://doi.org/10.1103/PhysRevD.99.034502}
  {\path{doi:10.1103/PhysRevD.99.034502}}.

\bibitem{Blossier:2009kd}
B.~Blossier, M.~Della~Morte, G.~von Hippel, T.~Mendes, R.~Sommer, {On the
  generalized eigenvalue method for energies and matrix elements in lattice
  field theory}, JHEP 04 (2009) 094.
\newblock \href {http://arxiv.org/abs/0902.1265} {\path{arXiv:0902.1265}},
  \href {https://doi.org/10.1088/1126-6708/2009/04/094}
  {\path{doi:10.1088/1126-6708/2009/04/094}}.

\bibitem{Dudek:2009qf}
J.~J. Dudek, R.~G. Edwards, M.~J. Peardon, D.~G. Richards, C.~E. Thomas,
  {Highly excited and exotic meson spectrum from dynamical lattice QCD}, Phys.
  Rev. Lett. 103 (2009) 262001.
\newblock \href {http://arxiv.org/abs/0909.0200} {\path{arXiv:0909.0200}},
  \href {https://doi.org/10.1103/PhysRevLett.103.262001}
  {\path{doi:10.1103/PhysRevLett.103.262001}}.

\bibitem{Dudek:2010wm}
J.~J. Dudek, R.~G. Edwards, M.~J. Peardon, D.~G. Richards, C.~E. Thomas,
  {Toward the excited meson spectrum of dynamical QCD}, Phys. Rev. D 82 (2010)
  034508.
\newblock \href {http://arxiv.org/abs/1004.4930} {\path{arXiv:1004.4930}},
  \href {https://doi.org/10.1103/PhysRevD.82.034508}
  {\path{doi:10.1103/PhysRevD.82.034508}}.

\bibitem{Bicudo:2021qxj}
P.~Bicudo, A.~Peters, S.~Velten, M.~Wagner, {Importance of meson-meson and of
  diquark-antidiquark creation operators for a $\bar{b} \bar{b} u d$
  tetraquark} (1 2021).
\newblock \href {http://arxiv.org/abs/2101.00723} {\path{arXiv:2101.00723}}.

\bibitem{Morningstar:1996ze}
C.~Morningstar, {Improved gluonic actions on anisotropic lattices}, Nucl. Phys.
  B Proc. Suppl. 53 (1997) 914--916.
\newblock \href {http://arxiv.org/abs/hep-lat/9608019}
  {\path{arXiv:hep-lat/9608019}}, \href
  {https://doi.org/10.1016/S0920-5632(96)00816-X}
  {\path{doi:10.1016/S0920-5632(96)00816-X}}.

\bibitem{Morningstar:website}
C.~Morningstar,
  \href{https://www.andrew.cmu.edu/user/cmorning/static_potentials/SU3_4D/B30_AR3/plots/Rplots.html}{Excitations
  of the static-quark potential, su(3) in 4 dimensions}.
\newline\urlprefix\url{https://www.andrew.cmu.edu/user/cmorning/static_potentials/SU3_4D/B30_AR3/plots/Rplots.html}

\bibitem{Brower:1981vt}
R.~Brower, P.~Rossi, C.-I. Tan, {The External Field Problem for {QCD}}, Nucl.
  Phys. B190 (1981) 699.
\newblock \href {https://doi.org/10.1016/0550-3213(81)90046-8}
  {\path{doi:10.1016/0550-3213(81)90046-8}}.

\bibitem{Parisi:1983hm}
G.~Parisi, R.~Petronzio, F.~Rapuano, {A Measurement of the String Tension Near
  the Continuum Limit}, Phys. Lett. B128 (1983) 418.
\newblock \href {https://doi.org/10.1016/0370-2693(83)90930-9}
  {\path{doi:10.1016/0370-2693(83)90930-9}}.

\bibitem{Albanese:1987ds}
M.~Albanese, et~al., {Glueball Masses and String Tension in Lattice QCD}, Phys.
  Lett. B 192 (1987) 163--169.
\newblock \href {https://doi.org/10.1016/0370-2693(87)91160-9}
  {\path{doi:10.1016/0370-2693(87)91160-9}}.

\bibitem{Morningstar:2003gk}
C.~Morningstar, M.~J. Peardon, {Analytic smearing of SU(3) link variables in
  lattice QCD}, Phys. Rev. D 69 (2004) 054501.
\newblock \href {http://arxiv.org/abs/hep-lat/0311018}
  {\path{arXiv:hep-lat/0311018}}, \href
  {https://doi.org/10.1103/PhysRevD.69.054501}
  {\path{doi:10.1103/PhysRevD.69.054501}}.

\bibitem{Edwards:1997xf}
R.~G. Edwards, U.~M. Heller, T.~R. Klassen, {Accurate scale determinations for
  the Wilson gauge action}, Nucl. Phys. B 517 (1998) 377--392.
\newblock \href {http://arxiv.org/abs/hep-lat/9711003}
  {\path{arXiv:hep-lat/9711003}}, \href
  {https://doi.org/10.1016/S0550-3213(98)80003-5}
  {\path{doi:10.1016/S0550-3213(98)80003-5}}.

\bibitem{Drummond:2002yg}
I.~T. Drummond, A.~Hart, R.~R. Horgan, L.~C. Storoni, {One loop calculation of
  the renormalized anisotropy for improved anisotropic gluon actions on a
  lattice}, Phys. Rev. D 66 (2002) 094509.
\newblock \href {http://arxiv.org/abs/hep-lat/0208010}
  {\path{arXiv:hep-lat/0208010}}, \href
  {https://doi.org/10.1103/PhysRevD.66.094509}
  {\path{doi:10.1103/PhysRevD.66.094509}}.

\bibitem{Drummond:2003qu}
I.~T. Drummond, A.~Hart, R.~R. Horgan, L.~C. Storoni, {The Contribution of
  O(alpha) radiative corrections to the renormalized anisotropy and application
  to general tadpole improvement schemes: Addendum to `One loop calculation of
  the renormalized anisotropy for improved anisotropic gluon actions on a
  lattice'}, Phys. Rev. D 68~(5) (2003) 057501.
\newblock \href {http://arxiv.org/abs/hep-lat/0307010}
  {\path{arXiv:hep-lat/0307010}}, \href
  {https://doi.org/10.1103/PhysRevD.68.057501}
  {\path{doi:10.1103/PhysRevD.68.057501}}.

\bibitem{Morningstar:website1}
Morningstar.
\newblock
  \href{https://www.andrew.cmu.edu/user/cmorning/static_potentials/SU3_4D/greet.html}{[link]}.
\newline\urlprefix\url{https://www.andrew.cmu.edu/user/cmorning/static_potentials/SU3_4D/greet.html}

\bibitem{Luscher:1980iy}
M.~Luscher, G.~Munster, P.~Weisz, {How Thick Are Chromoelectric Flux Tubes?},
  Nucl. Phys. B 180 (1981) 1--12.
\newblock \href {https://doi.org/10.1016/0550-3213(81)90151-6}
  {\path{doi:10.1016/0550-3213(81)90151-6}}.

\bibitem{Bali:2005fu}
G.~S. Bali, H.~Neff, T.~Duessel, T.~Lippert, K.~Schilling, {Observation of
  string breaking in QCD}, Phys. Rev. D 71 (2005) 114513.
\newblock \href {http://arxiv.org/abs/hep-lat/0505012}
  {\path{arXiv:hep-lat/0505012}}, \href
  {https://doi.org/10.1103/PhysRevD.71.114513}
  {\path{doi:10.1103/PhysRevD.71.114513}}.

\bibitem{Bulava:2019iut}
J.~Bulava, B.~H\"orz, F.~Knechtli, V.~Koch, G.~Moir, C.~Morningstar,
  M.~Peardon, {String breaking by light and strange quarks in QCD}, Phys. Lett.
  B 793 (2019) 493--498.
\newblock \href {http://arxiv.org/abs/1902.04006} {\path{arXiv:1902.04006}},
  \href {https://doi.org/10.1016/j.physletb.2019.05.018}
  {\path{doi:10.1016/j.physletb.2019.05.018}}.

\bibitem{Bicudo:2019ymo}
P.~Bicudo, M.~Cardoso, N.~Cardoso, M.~Wagner, {Bottomonium resonances with $I =
  0$ from lattice QCD correlation functions with static and light quarks},
  Phys. Rev. D 101~(3) (2020) 034503.
\newblock \href {http://arxiv.org/abs/1910.04827} {\path{arXiv:1910.04827}},
  \href {https://doi.org/10.1103/PhysRevD.101.034503}
  {\path{doi:10.1103/PhysRevD.101.034503}}.

\end{thebibliography}

%%%%%%%%%%%%%%%%%%%%%%%%%%%%%%%%%%%%%%%%%%%%%%%%%%%%%%%%%%%%%%%%%%%%%%%%%%%%%
\end{document}